\documentclass[preprint]{aastex631}

\usepackage{float}
\usepackage{bm}

\makeatletter
\let\original@singlespace\singlespace
\let\original@doublespace\doublespace
\makeatother

\usepackage{soul}

\usepackage{tcolorbox}
\tcbuselibrary{
    skins,         
    breakable,     
    theorems,      
    listings,      
    xparse         
}

\begin{document}

\title{A ``Rosetta Stone" for Studies of Spatial Variation in Astrophysical Data: \\ Power Spectra, Semivariograms, Structure Functions, and More}

\author[0000-0002-8632-6049]{Benjamin Metha}
\email{methab@student.unimelb.edu.au}
\affiliation{University of Melbourne, 1 Tin Alley, Parkville 3050 Victoria, Australia}
\affiliation{Australian Research Council Centre of Excellence for All-Sky Astrophysics in 3-Dimensions, Australia}
\author[0000-0002-4064-7883]{Sabrina Berger}
\thanks{Both SB and BM contributed equally to this work, and should be treated as corresponding authors and first-authors for citation purposes.}
\email{sabrina.berger@student.unimelb.edu.au}
\affiliation{University of Melbourne, 1 Tin Alley, Parkville 3050 Victoria, Australia}
\affiliation{Australian Research Council Centre of Excellence for All-Sky Astrophysics in 3-Dimensions, Australia}
\affiliation{Research School of Astronomy and Astrophysics, Australian National University, Canberra, ACT 2611, Australia}

\begin{abstract}
\noindent From the turbulent interstellar medium to the cosmic web, astronomers in many different fields 
have needed to make sense of spatial data describing our Universe. Through different historical choices for mathematical conventions, many different subfields of spatial data analysis have evolved their own language for analysing structures and quantifying correlation in spatial data.
Because of this history, terminology from a myriad of different fields is used, often to describe two data products that are mathematically identical.
In this Note, we define and describe the differences and similarities between the power spectrum, the two-point correlation function, the covariance function, the semivariogram, and the structure functions, in an effort to unify the languages used to study spatial correlation. We also highlight under which conditions these data products are useful and describe how the results found using one method can be translated to those found using another, allowing for easier comparison between different subfields' native methods. We hope for this document to be a ``Rosetta Stone" for translating between different statistical approaches, allowing results to be shared between researchers from different backgrounds, facilitating more cross-disciplinary approaches to data analysis.
\end{abstract}

\section{Introduction} 
\label{sec:intro}

\textit{Things that are close to each other tend to be similar in other ways.} This general principle, sometimes referred to as Tobler's First Law of Geography \citep{Tobler1970}, describes \emph{everything}. Hot days tend to follow hot days, and cold days tend to follow cold ones. People who live in the same area tend to vote similarly, earn a similar amount of money, drive similar cars, and live to about the same age. On the smallest scales, mosquitoes that are captured from the same local environment are more genetically similar than those that are taken from separate locations. The concentration of malaria in their bloodstreams will be more similar if they are drawn from nearby locations, as will the level of drug resistance within the malaria parasites. On the largest scales, the structure of the Universe also follows Tobler's First Law: regions that are rich in matter tend to be close to other dense regions of the Cosmos. 

It is of no surprise, then, that mathematicians from many diverse disciplines have made attempts to capture the ways that the similarity between different things depends on their distance. In its theoretical form, mathematics is a language that precisely and unambiguously describes the Universe. It is both invented \emph{and} discovered: we invent words to describe the things we see, and discover relationships between them.

As with any language, mathematics has dialects. As these different areas of study evolved, all of them independently tried to solve the problem of how to best describe the ways in which nearby things resemble each other. Like finches isolated on separate islands of the Galapagos, over time, each subfield organically developed its own set of methods, tools, and techniques, each of which capture the same kind of characteristics about a spatial data set.\footnote{Indeed, one could use Tobler's First Law of Geography to describe this kind of mathematical-linguistic speciation: scientific disciplines that are close to each other (in terms of both the kinds of things that they study and where and when these investigations historically happened) tend to follow similar statistical naming conventions.} 

Today, researchers who come from separate ``islands of study" have great difficulty understanding one another. The investigator who wishes to read widely, connect to other cultures, and discover cross-disciplinary approaches to analyse their data must sail through treacherous waters, infested with scary mathematical functions such as \textit{variograms} and \textit{semivariograms}; the \textit{two-point correlation function}, \textit{autocorrelation function}, \textit{autocovariance function} and \textit{cross-correlation function}; the \textit{power spectrum} and \textit{power spectral density}; the \textit{energy spectrum}; and \textit{structure functions} of the first, second, and higher orders. This is a shame, as it is often the case that a problem in one field has already been solved by a mathematical approach that is native to a different field -- but unless we can talk to each other, there is no way that we can learn from each other.

The purpose of this document is to help solve this problem. We are two graduate students in astronomy who have been taught how to analyse spatial data on two different mathematical-linguistic islands (Benjamin Metha speaks the language of geostatistics, Section \ref{ssec:geostats}; and Sabrina Berger's native tongue is the Fourier method of power spectrum analysis, Section \ref{ssec:cosmo}). In an attempt to understand each others' methods, we have crossed a dense jungle of nomenclature. We leave this Note behind as a collection of terminological trail markers, so that others can traverse the rough seas between each island with ease.
Our intended reader is both the new student and the expert in one of these particular methods, who is curious about the similarities and differences between their approach and other methods that have been tried and tested in the literature. We hope that this piece serves as a ``Rosetta stone," allowing methods from one discipline to be translated into the language of another, enabling results to be shared between previously-isolated communities, and facilitating cross-disciplinary collaboration.

Before we delve into the myriad of languages that are used to describe spatial data, it helps to have a common language that we can speak. In Section \ref{sec:classical_stats}, we present some important tools from classical statistics that capture the broad properties of random variables: the \textit{mean} (what value is a random variable on average?), \textit{variance} (how similar do observed values of the random variable tend to be?), \textit{covariance} (how much do changes in this random variable tend to imply changes in another random variable?), and \textit{correlation} (a normalised version of the covariance that shows how much information a random variable contains about another). Once we have built a solid bedrock, we then extend these definitions to work on \textit{random fields} (random variables that are observed over a spatial domain) and \textit{time series data} (random variables that are observed over time) in Section \ref{sec:spatial_stats}. While this content may be familiar to a large fraction of our audience, we have discovered that there is still a tangle of terminology and some disagreement on definitions in these shallow waters. We encourage even the seasoned statistician to at least skim this Section, pausing to ensure that the definitions they are familiar with match our own.

We then embark on an exploration into the way that spatial data is analysed in three different fields of study. We begin in Section \ref{ssec:geostats} to cover how geostatisticians use the \textit{semivariogram} to characterise correlation in spatially-varying data. In Section \ref{ssec:cosmo}, we leave the real world behind and travel into Fourier space to explore how cosmologists use the \textit{power spectrum} to extract the same characteristics. Next, in Section \ref{ssec:turbulence}, we visit the chaotic realm of turbulence. We learn how two tools used by fluid dynamicists to capture structure in stochastic environments, the \textit{energy spectrum} and the \textit{structure functions}, are related to the other approaches that we have encountered. We reflect on our journey in Sections \ref{sec:reflections} and \ref{sec:summary}, drawing bridges between these different domains, and discussing which data products each approach is best suited for, and how results found with one approach are connected to the other approaches. 
Finally, we provide a glossary that gives a plain English definition 
to all of the mathematical terms that we have encountered in this Note, to assist the curious researcher who wants to read more about the approaches to spatial data analysis used in other fields.

To accompany this pedagogical article, we have also created an interactive Jupyter notebook\footnote{Available on \href{https://github.com/sabrinastronomy/rosetta-stone}{SB's GitHub}.} that contains Python implementations of every method that we cover on this tour, and a selection of random fields on which they can be applied. We hope that this is useful for the reader (i) to gain some hands-on intuition on how these methods work numerically and what information they tell you, and (ii) to have some off-the-shelf implementations of these methods available that you (yes, you!) can use for your own research.

To keep this Note relatively brief, we limit our investigations to methods that capture only the first- and second-order structure of a random field.\footnote{This information is sufficient to completely describe \textit{Gaussian} random fields. For other kinds of random fields, higher-order statistics are needed to capture deviations from Gaussianity -- but that's another story. Even if your field is not Gaussian, you can still learn a lot about how it behaves from its first and second-order statistics.} For this reason, some words that will not appear in this manuscript that describe related ways of capturing spatial correlation\footnote{Except for right here, and in the Glossary.} include the \textit{bispectrum}, the \textit{trispectrum}, the \textit{three-point correlation function}, the \textit{scattering transform} (First applied to cosmology in \citet{2020MNRAS.499.5902C}), nor any discussion of machine and deep learning (e.g. convolutional neural networks).

\label{sec:warmup}
\section{Fundamentals}
\label{sec:classical_stats}

When we make measurements in the real world, the data that we acquire are always uncertain, and the processes that produce the data that we see often contain an element of randomness. In order to make sense of what we see, it is important to understand how much we don't know about what we observe. To understand the external, uncertain world that we see around us, we must use \textit{statistics}. To capture this uncertainty without losing any information, statisticians define the things that we observe in terms of \emph{random variables}.

A random variable is a mathematical way of representing outcomes that you are not sure about. When you try to measure the value of a random variable, there are a range of different possible answers that you could get. Ask a person on the street their age, and you could feasibly get any number between two and one hundred.\footnote{Note: we do not recommend that you run this experiment in real life, as it is rude to ask strangers their age.} If you repeat this experiment with a different person, you will probably get a different number. We use capital letters ($X,Y$) to denote these random variables, and lowercase letters with subscripts to denote the different values that we measure for them. In this scenario, let $X$ be the random variable ``the age of a person on this street". Then $x_1$ is the age that the first person that we ask tells us, $x_2$ is the age that the second person tells us, and so on. If we were to also ask each person for their height, then we could let $Y$ be the random variable ``the height of a person on this street", and this random variable would have the sampled values $y_1, y_2$, and so on for each person.

When we analyse random variables, our goal is to get some idea of their \textit{distribution} and how they depend on other variables to inform our understanding about what's going on. The probability distribution of a random variable tells you all of the possible values that the random variable can take, and how likely each of them is to occur. By continually taking many samples of a random variable (asking lots of people on the street their age, in our example), we get some idea of what the \textit{typical} values of a random variable are, and how much they tend to \textit{vary}. If we were to conduct our experiments just outside a primary school, we would get a different typical value for the ages of people than we would if we tried this experiment in a jazz bar. On the other hand, if we compared the ages of a sample of random people in a primary school to the ages of a sample of random people in a third-grade maths classroom, we might get similar typical values for both random variables, but we would find that the values that we measure tend to vary a lot more in the first case than in the second one. We make these concepts more formal below.

\subsection{Measures of centre and measures of spread: mean, variance, and standard deviation}
\label{sec:average}
Let $X$ be a random variable that we have measured $n$ times, with values of $x_1, x_2, \dots, x_n$. We want to know two things about the distribution of this random variable. Firstly, what is the value of this random variable \textit{roughly} (mean)? Secondly, how similar do observed values of the random variable tend
to be (variance)\footnote{In the world of probability, answering these questions is equivalent to estimating the first (raw) and second (central) \textit{moments} of the random variable's probability distribution function. You can find a definition of moments, their close cousins the \textit{cumulants}, and the way that they are related \href{https://ocw.mit.edu/courses/18-366-random-walks-and-diffusion-fall-2006/52f5cd1ad7315858f2759bdc2636ba0e_lec02.pdf}{here}.}? 

To answer our first question, we define the \textit{sample mean} to be:
\begin{equation}
    \label{eq:mean}
    \bar{x} = \widehat{\mathbb{E}[x]} = \frac{1}{n} \sum_{i=1}^{n} x_i.
\end{equation}
This value is an estimator\footnote{In this Note, we use estimator to mean a statistical estimator which uses a rule, such as Equation \ref{eq:mean} to \textit{estimate} a quantity -- see the glossary for further details.} for the \textit{expectation value} (or expected value) of a random variable. The expectation value of a random variable is the average value across all values that appear in its population.
By population, we mean the set of all possible observations of a random variable. The frequency of each individual value it can take is the probability of that event (defined below in both the discrete and continuous case). The set of each probability among possible outcomes comprises the random variable's \textit{probability distribution}. If we have \textit{samples} of a random variable, we can use Equation \ref{eq:mean} to estimate the expected value of a random variable. However, we could never hope to get our hands on the exact expectation value of a random variable from a finite number of measurements alone (if we're sampling from an infinite population). Instead, we would need to know the probability distribution of the random variable to calculate it exactly.

The expectation value of a discrete random variable, $X$, is:
\begin{equation}
    \label{eq:exp_value_theory_discrete}
    \mathbb{E}[X] \equiv \sum_{i=1}^{\infty} x_{i} p_{i}, 
\end{equation}
where $p_{i}$ is the probability mass function, or the discrete probability distribution function. That is, each value of $p_{i}$ is the probability that we will observe $X$ to have a value of $x_i$ (in mathematical language, this is written as $p_{i} = P(X=x_{i})$), and $x_i$ are all of the possible outcomes of $X$.

Sometimes, there are too many values that a random variable can take (for example, if the random variable in question can take any real number as a value), and this summation definition of the expectation value does not work.
For times like this, the expectation value of a random variable can also be defined as follows: 
\begin{equation}
    \label{eq:exp_value_theory_continuous}
    \mathbb{E}[X] \equiv \int_{-\infty}^{\infty} x p(x) dx,  
\end{equation}
where $p(x)$ is the \textit{probability density function} of our random variable $X$. The values of $p(x)$ are defined so that the integral of $p(x)$ between $x_1$ and $x_2$ is the probability that when we observe $X$, it lies between $x_1$ and $x_2$. In mathematical language, this is written as $P(x_{1}\leq X \leq x_2) = \int_{x_1}^{x_2} p(x)dx$. 
 
In Equations \ref{eq:exp_value_theory_discrete} and \ref{eq:exp_value_theory_continuous}, we defined the expectation value, $\mathbb{E}[X]$, which we cannot get exactly right unless we know the exact distribution the random variable $X$ was drawn from. The most common estimator is the sample mean (which we will denote by $\bar{x}$ in this Note) which is the average computed from a finite number of observations of $X$ (Equation \ref{eq:mean}).

To answer our second question (how much does this random variable tend to \textit{vary}?), we measure the spread of a random variable by calculating the \textit{variance} of our sample. As we did with \textit{mean}, we present the true definition of the variance as well as a common estimator. To compute the true variance, we calculate the expected value of the squared difference between the values of $X$ that we could measure, and their expected value $\mathbb{E}(X)$:
\begin{equation}
    \label{eq:var_theory}
     \mathrm{Var}[X] \equiv \mathbb{E}[\left(X-\mathbb{E}(X)\right)^2].
\end{equation} In this definition, $\mathrm{Var}[X]$  depends on the distribution of our random variable, and not on just a few samples that we observe -- that is, we need to know the distribution of the random variable to compute the variance, or at least the first and second moments. Since we hardly ever know the full distribution of a random variable in practice, we like to estimate the variance with the \textit{unbiased sample} variance. This is computed by taking the average squared difference of each data point from the sample mean:
\begin{equation}
    \label{eq:var}
    \mathrm{\widehat{Var}}(X) = \ \frac{1}{n-1} \sum_{i=1}^{n} (x_i - \bar{x})^2,
\end{equation}
where the wide hat denotes that this is a variance \textit{estimator}.

In both of these definitions, the units of variance will be the square of the units of the original variable.
For this reason, it is often convenient to talk about the \textit{standard deviation}, defined to be the square root of the variance, to describe how much the random variable tends to randomly vary. Because it has the same units as the original thing that was measured, they are easy to compare. Often, the symbol $\sigma$ (sigma) is used to denote the standard deviation.

\begin{tcolorbox}[colback=red!5!white,colframe=red!75!black]{
  \textbf{Wait, why are we dividing by $n-1$ and not $n$?} \\ 
  Technically, the above Equation (\ref{eq:var}) is not \textit{exactly} an arithmetic average, because we divide by $n-1$ and not $n$. If we have only samples from a population of a random variable, we only know the sample mean and not the true expectation value. If we know the expectation value, $\mathbb{E}(X)$, we can determine the true variance as $\rm{Var}(X) =\frac{1}{n} \sum_{i=1}^{n} (x_i - \mathbb{E}(X))^2$. We usually don't know $\mathbb{E}(X)$, though! If we calculate the sample variance using our best estimate of the sample mean as $ \widehat{\rm{Var}}(X) =\frac{1}{n} \sum_{i=1}^{n} (x_i - \bar{x})^2$, the result will be biased to be slightly lower than the true value of the variance.. Using $n-1$ instead of $n$ is called \href{https://en.wikipedia.org/wiki/Bessel%27s_correction}{Bessel's correction}, and it corrects for this bias, allowing us to estimate the true variance in an unbiased way. 
  
  As $n$ gets larger, the difference between using $n-1$ and $n$ becomes negligible, and in the limit where we have perfect knowledge of our random variable (as $n\to \infty$), the definitions that use $n$ and $n-1$ become exactly the same. For more intuition as to why dividing by $n-1$ is the right thing to do, watch this \href{https://www.youtube.com/watch?v=E3_408q1mjo}{6 minute video} (or 3 minutes at 2x speed) which walks you through an explicit example showing that this is true. Alternatively, a full proof is presented \href{https://gregorygundersen.com/blog/2019/01/11/bessel/}{here}.}
\end{tcolorbox}

The standard deviation has been very well studied, so we know a lot about how it is expected to behave, making it a very useful statistic. 
Just by knowing the mean and standard deviation of a random variable, we get a lot of information about its distribution. To show you what we mean, let's return to our example where $x$ is the age of a randomly-selected person on the street. After interviewing a good number of people, we calculate the sample mean of this random variable to be $\bar{x} = 40$ years and compute its sample variance to be $\rm \widehat{Var}(X) = 100$ years squared, giving us a standard deviation of $10$ years. From this information and the incorrect assumption that age follows a normal distribution (it doesn't because age can't be negative!), and our knowledge of how a standard deviation works, we can predict that about 68\% of the people on this street will be between 30 and 50 years of age (less than one standard deviation from the mean), about 95\% will be between 20 and 60 (less than two standard deviations away), and about 99.7\% will be between 10 and 70 (less than three standard deviations away). If the standard deviation was 5 instead, all of these age intervals would only be half as wide. If the standard deviation was 0, then this would mean that there is no variation in our data, and everyone on the street is exactly the same age.

\subsection{Defining correlation for random variables}
\label{sec:corr_cov}
Now, let's consider the case where we have two random variables, $X$ and $Y$ -- say, age and height of random people in a population. We take a different measurement of both $X$ and $Y$ for each of $n$ people, giving us samples with values of $(x_1, y_1), (x_2, y_2), \dots, (x_n, y_n)$. We want to find a statistic that can answer the following question: how related are these two random variables? Does knowing about one of them give you any information about the other -- that is, can you use someone's height to predict their age, or vice versa? 

When the relationship between X and Y is monotonic, the word for what we are trying to measure is \textit{correlation}. If $X$ and $Y$ are positively correlated, then if $X$ is measured to have a value that is higher than its mean, then $Y$ will likely be higher than its mean, too. If they are negatively correlated, then if $X$ is measured to be higher than its mean, $Y$ is (on average) lower than its mean. The final option is that these two random variables are \textit{independent}: measuring $X$ does not give us any additional information about $Y$, and measuring $Y$ does not give us any additional information about $X$.

As a first attempt to capture this information about this pair of variables, we define the \textit{covariance} between $X$ and $Y$ as:
\begin{equation}
    \label{eq:cov_theory}
    \text{Cov}(X,Y) \equiv \mathbb{E}[(X - \mathbb{E}[X])(Y - \mathbb{E}[Y])].
\end{equation}
However, we can't usually know the expectation values of our random variables \textit{exactly}. So in practice, we instead estimate the covariance with the unbiased sample covariance as follows:\footnote{Note the $\frac{1}{n-1}$ in the denominator; this is Bessel's correction, and it works to unbias the covariance in the same way that it unbiases the variance estimated in Equation \ref{eq:var}.} 
\begin{equation}
    \label{eq:cov}
    \widehat{\rm {Cov}}(X,Y) = \frac{1}{n-1} \sum_{i=1}^{n} (x_i - \bar{x})(y_i - \bar{y}).
\end{equation}

This statistic does a good job of capturing the information that we are interested in. 
If $X$ tends to rise when $Y$ rises, and tends to fall when $Y$ falls, then the values inside the summand
will tend to be positive -- and $\widehat{\rm{Cov}}(X,Y)$, on the whole, will be a positive number. The more often this happens, the higher the value of $\widehat{\rm{Cov}}(X,Y)$ -- so $\widehat{\rm{Cov}}(X,Y)$ is sensitive to the level of correlation between these two variables. On the other hand, if $X$ tends to fall when $Y$ rises and rise when $Y$ falls, then $\widehat{\rm{Cov}}(X,Y)$ will more often than not be negative, so $\widehat{\rm{Cov}}(X,Y)$ will be negative overall. On an unexpected third hand, if there is no connection between $X$ and $Y$, then the product $(x_i - \bar{x})(y_i - \bar{y})$ will be equally likely to be positive or negative for each pair $(x_i, y_i)$ -- so after averaging over all pairs, the covariance should be close to zero.

The problem with using covariance to measure the similarity between two variables is that it is difficult to interpret. Firstly, it has weird units -- $\widehat{\rm{Cov}}(X,Y)$ has the units of X multiplied by the units of Y. If $X$ is age in years and $Y$ is height in feet, then $\widehat{\rm{Cov}}(X,Y)$ will have the unusual units of foot-years. If we measured a covariance value of 0.7 foot-years for these two random variables, would you think that they are more correlated or less correlated than you expected? 

Secondly, the covariance does not only depend on the correlation between these two variables, but also on how much they vary individually. If the variance on $X$ is larger, then the values of $(x_i - \bar{x})$ will be larger, too -- so the covariance will rise. In practise, this means that if we measured the heights and ages of babies over a six month period and calculated the covariance, we would get a number that is about four times smaller than we would if we tried the same experiment using a years' worth of data. At first glance, an age-height covariance of 0.06 foot-years seems to imply that two variables are less strongly connected than if a pair of random variables had a covariance of 0.2 foot-years, but unless we know what the variances of both $X$ and $Y$ are, we cannot say this for sure. 

All things considered, the covariance could be more useful if it were normalised by the amplitudes of the variances in a nice way that would rid of all the weird units. So that's exactly what we do. We define the \textit{correlation} ($\rho$) between X and Y to be the covariance between them, normalised by their standard deviations (the square root of their variances):

\begin{equation}
    \label{eq:corr_perfect}
    \rho(X,Y) = \frac{\rm{Cov}(X,Y)}{\sqrt{\rm {Var}(X)\rm {Var}(Y)}},
\end{equation}
or the sample correlation version:
\begin{equation}
    \label{eq:corr}
    \widehat{\rho}(X,Y) = \frac{\widehat{\rm{Cov}}(X,Y)}{\sqrt{\rm \widehat{Var}(X)\rm \widehat{Var}(Y)}}.
\end{equation}

The definition of correlation in equation \ref{eq:corr} 
(often called the Pearson correlation coefficient\footnote{Named after Karl Pearson, who stole it from Francis Galton, who invented it all by himself -- about 45 years after it had been invented independently by French mathematician Auguste Bravais \citep{Bravais44}.}) has some nice mathematical properties. For perfectly correlated data, $\rho = 1$. You can see this for yourself by trying to compute $\rho(X,X)$ (because $X$ is perfectly correlated with itself) and simplifying:
\begin{equation}
    \rho(X,X) = \frac{\widehat{\text{Cov}}(X,X)}{\sqrt{\widehat{\text{Var}}(X)\widehat{\text{Var}(X)}}} = 
    \frac{\widehat{\text{Var}}(X)}{\widehat{\text{Var}}(X)} = 1.
\end{equation}
Similarly, for perfectly anticorrelated data (like $X$ and $-X$), $\rho = -1$. For independent $X$ and $Y$, $\rho(X,Y)$ will be zero, because the covariance between $X$ and $Y$ will be zero.
The value of $\rho$, then, can be thought of as a statistic that tells you (i) whether two variables are positively or negatively correlated, and (ii) how strongly related they are on a scale of ``not at all" ($\rho=0$) to ``completely" ($\rho= \pm 1$).\footnote{
Note that Pearson's correlation coefficient only captures \textit{linear} correlations between variables. This can be a problem if two variables depend on each other in a nonlinear way. A cute example that you can build at home is the random variable pair where $Y=\sin(X)$, and $X$ ranges from $0$ to $20$. If you know $X$, then you can predict $Y$ perfectly, but if you calculate the Pearson correlation coefficient between $X$ and $Y$, it will be close to zero. Other kinds of correlation coefficients have been invented to study non-linear correlations between random variables. See this nice \href{https://goldinlocks.github.io/Comparing-Linear-and-Nonlinear-Correlations/}{notebook} for an overview.} 

\section{Extending these definitions for spatial statistics}
\label{sec:spatial_stats}

Now that we have defined covariance and correlation for random variables, 
we are ready to discuss how to naturally extend these definitions to learn about correlation within random fields. In doing so, we will be introducing a lot of terminology. We warn the reader who has some familiarity with this area to proceed with caution. Many of these functions have been given different names in different fields. We have chosen our definitions and notation as carefully as possible to serve as natural extensions of the statistical definitions defined in the previous Section that everyone agrees upon.

We define a random field over a domain $D$ such that at each point $\vec{x}$ in our domain, the value of $Z(\vec{x})$ is a random variable. If we observe $Z$ at $n$ points within our domain, then we get a sample of this random field, with values $Z(\vec{x_1}), Z(\vec{x_2}), \dots, Z(\vec{x_n})$. For the purposes of this document, we assume that each $Z(\vec{x_i})$ is a real number.

The naive statistical approach to analysing this type of data is to \textit{forget the location that each data point was drawn from}, and work with our data treating it like we have samples of a random variable $Z_1, Z_2, \dots, Z_n$. We could then use the tricks we learned in Section \ref{sec:classical_stats} to estimate the mean (Equation \ref{eq:mean}) and the variance (Equation \ref{eq:var}) of our random field using this sample. However, if we do it this way, we are losing a lot of information. 

\begin{figure}
    \centering
    \includegraphics[width=0.95\textwidth]{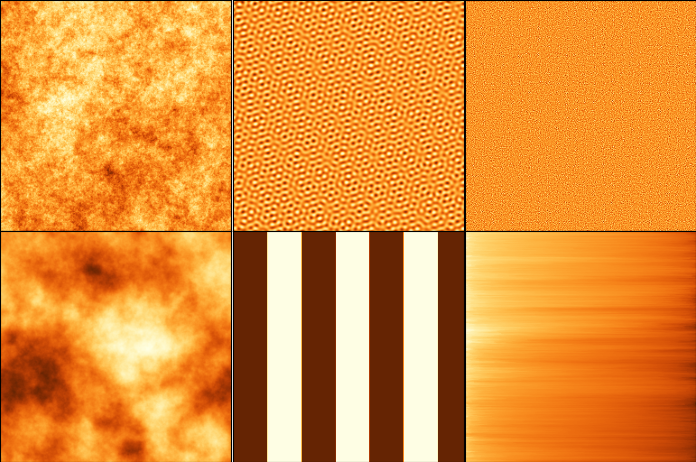}
    \caption{Six different random fields. All of these fields were constructed to have the same means and the same variance, but they look (and are) wildly different. In order to classify and quantify how these field are different, we need to consider the spatial aspects of our data. The reader can see how we generated these fields by taking a look at the corresponding  \href{https://github.com/sabrinastronomy/rosetta-stone}{Jupyter Notebook tutorial} for this note.}
    \label{fig:anscombe}
\end{figure}

To illustrate this point, in Figure \ref{fig:anscombe}, we show six different random fields. All of these random
fields were constructed to have exactly the same mean and variance.
Because of this, if you were
to take the standard statistical approach of forgetting about the locations of each pixel, you would not be able to tell the difference between any of the distributions shown below. However, a quick visual inspection\footnote{This is the formal scientific way of saying ``just look at it!"} is enough to tell you that all of these random fields are very different. Some are highly ordered, with similar values of $Z(\vec{x})$ being consistently found in nearby spatial locations. Some are highly chaotic, showing no clear structure at all. Some evolve much more rapidly over shorter spatial scales than others, whereas others remain correlated over larger distances. When we forget about the spatial location $\vec{x}_i$ that each sample of $Z$ is drawn from, we lose all of this information. Throughout the 20$^\text{th}$ century, mathematicians from all over the world working in many different fields noticed this, and all came to the same conclusion: we can do better. This led to the invention of a number of methodologies, all with the same end-goal in mind: to \textit{quantify spatial correlations}.

\subsection{Defining correlation for random fields
\footnote{or: Wait, why are B.M. and S.B. suddenly fighting?}
}
\label{ssec:random_field_correlation}
Looking at the data fields above, we can see that for at least some of them, nearby data points appear to be \textit{correlated}. That is, values of $Z$ that are drawn from points that are close to each other tend to have similar values. In other words, some of these fields seem to obey Tobler's First Law of Geography. If we use the methods of estimating mean and variance described in Section \ref{sec:classical_stats}, we'll only be left with two numbers to describe each of these random fields. This is insufficient to describe the complexities of these subjects. We want to find a way that we can compute correlations between values of $Z(\vec{x})$ in the same random field that come from different spatial locations. Doing this kind of analysis requires us to know the location ($\vec{x}$) that each measurement ($Z(\vec{x})$) is taken from.

The most fundamental way to quantify this relation is to treat every measurement at each location as being generated by a different, separate random variable. We can then look to see if there are any relationships between any pair of measurements. To do this, we define the \textit{covariance matrix} as the \textit{symmetric}, \textit{square} matrix whose $i,j$-th element is the covariance between $Z(\vec{x_i})$ and $Z(\vec{x_j})$:

\begin{equation}
    \label{eq:cov_matrix}
     \widehat{\mathbf{Cov(Z)}} = \begin{bmatrix}
 \widehat{\rm Cov} (Z(\vec{x}_1), Z(\vec{x}_1)) &  \widehat{\rm Cov} (Z(\vec{x}_1), Z(\vec{x}_2))  & \cdots &  \widehat{\rm Cov} (Z(\vec{x}_1), Z(\vec{x}_n))   \\
 \widehat{\rm Cov} (Z(\vec{x}_2), Z(\vec{x}_1)) &  \widehat{\rm Cov} (Z(\vec{x}_2), Z(\vec{x}_2))  & \cdots &  \widehat{\rm Cov} (Z(\vec{x}_2), Z(\vec{x}_n))   \\
\vdots & \vdots & \ddots & \vdots \\
 \widehat{\rm Cov} (Z(\vec{x}_n), Z(\vec{x}_1)) &  \widehat{\rm Cov} (Z(\vec{x}_n), Z(\vec{x}_2))  & \cdots &  \widehat{\rm Cov} ( Z(\vec{x}_n), Z(\vec{x}_n))   \\
\end{bmatrix}
\end{equation}
Once we know the covariance matrix of our random field, we can then divide it by the estimated variance of our random field (a real number which we can compute using standard statistical methods) to produce the \textit{correlation matrix} -- the matrix whose $i,j$-th element is the Pearson correlation coefficient between $Z(\vec{x_i})$ and $Z(\vec{x_j})$:

\begin{equation}
    \label{eq:corr_matrix}
    \bm{\hat{\rho}(Z)} = \begin{bmatrix}
\hat{\rho}(Z(\vec{x}_1), Z(\vec{x}_1)) & \hat{\rho}(Z(\vec{x}_1), Z(\vec{x}_2))  & \cdots & \hat{\rho}(Z(\vec{x}_1), Z(\vec{x}_n))   \\
\hat{\rho}(Z(\vec{x}_2), Z(\vec{x}_1)) & \hat{\rho}(Z(\vec{x}_2), Z(\vec{x}_2))  & \cdots & \hat{\rho}(Z(\vec{x}_2), Z(\vec{x}_n))   \\
\vdots & \vdots & \ddots & \vdots \\
\hat{\rho}(Z(\vec{x}_n), Z(\vec{x}_1)) &\hat{\rho}(Z(\vec{x}_n), Z(\vec{x}_2))  & \cdots & \hat{\rho}(Z(\vec{x}_n), Z(\vec{x}_n))   \\
\end{bmatrix}
\end{equation}

This seems useful. However, this construction comes with an irritating caveat. Calculating the covariance matrix requires multiple measurements of the same locations in our field. When we just have one measurement, our entire covariance matrix is undefined and cannot be estimated, since we would need to divide by zero in Equation \ref{eq:cov}.\footnote{If, instead, we had a distribution of values of $Z(\vec{x}_i)$ at each location $\vec{x}_i$, or a way to estimate the distribution of $Z(\vec{x}_i)$ at each location $\vec{x_i}$, we would be fine. In these scenarios, covariance and correlation matrices are perfectly reasonable things to compute. The Python package \texttt{numpy} contains methods to compute a covariance matrix (\href{https://numpy.org/doc/stable/reference/generated/numpy.cov.html}{\texttt{numpy.cov}}) and a correlation matrix (\href{https://numpy.org/doc/stable/reference/generated/numpy.corrcoef.html}{\texttt{numpy.corrcoef}}) from an array of $N$ samples of $M$ random variables. }

But what if we just had a single measurement of $Z(\vec{x})$ at each location $\vec{x}$ and we wanted to quantify spatial correlation? Fortunately, there is a way forward -- but it relies on us knowing (or at least assuming) something about how our random fields behave. Looking at the first four
random fields in Figure \ref{fig:anscombe}, we can see that they all have two things in common. Firstly, on large scales, the mean value of the data does not seem to vary with space -- that is , that there is no global trend of the mean value of the data in these fields changing over scales that are the size of the data field, as is seen in the bottom-right panel of Figure \ref{fig:anscombe}. Secondly, the covariance between data at points $\vec{x}$ and $\vec{y}$ does not seem to depend on their absolute locations, but only on their separation. In mathematical language, this is the same as saying that at any two points $\vec{x}$ and $\vec{y}$, given any separation vector $\vec{r}$:
\begin{eqnarray}
\label{eq:weak-sense-stationary}
\mathbb{E}[Z(\vec{x})] &=& \mathbb{E}[Z(\vec{y})], \nonumber \\
&\text{and}& \\
    \text{Cov}[Z(\vec{x} + \vec{r}), Z(\vec{x})] & = &\text{Cov}[Z(\vec{y} + \vec{r}), Z(\vec{y})].\nonumber
\end{eqnarray}

There are many different names for this condition. In the arena of signal-processing (where $\vec{x}$ is usually a one-dimensional vector representing time), a random field (or time-varying signal) that follows these two conditions is called \textit{weakly stationary}, \textit{weak-sense stationary}, \textit{wide-sense stationary}, or \textit{second-order stationary}. These terms have been adopted to describe 2+ dimensional data in the geostatistical literature -- we will follow this convention and adopt the term \textit{second-order stationary} to describe these fields for the rest of this Note. Some people choose to simply call data that follows this condition \textit{stationary}, but people will often also use this word for a much stronger condition.\footnote{Stationary is also used to mean \textit{strictly stationary}, or \textit{strongly stationary}. Under this condition, all higher-order moments also depend only on the separation between data points and not their positions -- but that's beyond the scope of this work.} Another term used for data that follows this condition (for example, in cosmology) is to say that it is \textit{translationally-invariant}, or \textit{homogeneous} -- but similar to the word \textit{stationary}, these words are also sometimes used to mean a different, stronger condition.\footnote{In standard cosmologies, the Universe is assumed to be \textit{homogeneous} in the strong sense on large scales -- that is, all higher order moments are assumed be statistically the same for all pairs, or triples, or quadruples of points that are separated by the same distances, irrespective of their positions. Just as it does for the power spectrum with second-order homogeneity, this assumption allows higher-order statistics, such as the \textit{bispectrum} and the \textit{trispectrum} of the cosmic microwave background, to be computed from a single realisation of the data.}

No matter what you call it, if the random field $Z(\vec{x})$ follows this condition (\ref{eq:weak-sense-stationary}), then the covariance between two data points depends only on their separation. Because of this, we can define the \textit{covariance function} $C(\vec{r})$ to be the function that gives the covariance between anything separated by $\vec{r}$:
\begin{equation}
    \label{eq:cov_fn}
    C(\vec{r}) =  \widehat{\rm Cov}(Z(\vec{x} + \vec{r}), Z(\vec{x})).
\end{equation}

This time, the average in the expression for Cov (Equation \ref{eq:cov}) is computed over all pairs of points separated by $\vec{r}$ (or in practise, separated by $\vec{r}$). Since there is more than one point separated by each $\vec{r}$, there is no problem computing this for most values of $\vec{r}$.

Once we have this function, then we can divide it by the variance of our random field to get the Pearson correlation coefficient between any pair of data points separated by $\vec{r}$. Because this is the most natural extension of the definition of correlation for random variables, we call this function the \textit{correlation function}:
\begin{equation}
    \label{eq:corr_fn}
    \rho(\vec{r}) = \frac{C(\vec{r})}{\widehat{\rm Var}[Z(\vec{x})]} = \rho[ Z(\vec{x} + \vec{r}), Z(\vec{x})].
\end{equation}
    
If our random fields are very well-behaved, we can  simplify these functions one step further. Returning to Figure \ref{fig:anscombe}, we can see that the random fields that we show in our first four panels are \textit{isotropic}, or \textit{rotationally-invariant}: that is, there is \textit{no preferred direction} along which the data seems to be varying any more or less than in any other direction (the last two random fields in Figure \ref{fig:anscombe} do not have this property). If this is the case, then the covariance between two data points that are separated by $\vec{r}$ will depend only on the magnitude of $\vec{r}$. Under this condition, our covariance and correlation functions simplify to:

\begin{equation}
    C(r) = \widehat{\text{Cov}}(Z(\vec{x}), Z(\vec{y}))  \label{eq:isotropic_cov_fn} 
\end{equation}

\begin{equation}
\rho(r) = \rho( Z(\vec{x}), Z(\vec{y})) \label{eq:isotropic_corr_fn}
\end{equation}
where this time, the averages used to calculate the covariance (Equation \ref{eq:cov}) are taken over all pairs of points $\vec{x}$ and $\vec{y}$ for which $|\vec{x}-\vec{y}| = r$.
\newpage
\begin{tcolorbox}[colback=red!5!white,colframe=red!75!black]{
  \textbf{Wait, isn't this the two-point correlation function?} \\ 
  As this is a Note geared towards astronomers and astrophysicists, it 
  would be remiss of us to not mention the two-point correlation function.\footnote{Don't worry if you haven't heard of it. \href{https://www.quora.com/Are-the-2-point-correlation-function-and-the-autocorrelation-function-the-same}{A data scientist with a PhD in statistics had not heard of it, \newline either}. Seemingly, this function is seldom seen outside of astronomy.} Unfortunately, several different definitions for this function are used, but the most commonly-used one is this: for a real-valued random field $Z(\vec{x})$, the two-point correlation $\xi$ is defined to be:
\begin{equation}
\label{eq:2pcf}
    \xi(\vec{x}, \vec{y}) = \mathbb{E} \left[ Z(\vec{x}) Z(\vec{y}) \right]
\end{equation}}
If the random field $Z(\vec{x})$ is second-order stationary (\ref{eq:weak-sense-stationary} is true for all points $\vec{x}$ and $\vec{y}$), then $\xi$ depends only on the separation between data points:
\begin{equation}
\label{eq:2pcf_homogeneous}
    \xi(\vec{r}) = \mathbb{E} \left[ Z(\vec{x}) Z(\vec{y}) \right], \quad \text{where} \quad \vec{r} = \vec{x} - \vec{y}.
\end{equation}
And if $Z(\vec{x})$ is also isotropic, then $\xi$ depends only on the distance between data points:
\begin{equation}
\label{eq:2pcf_homogeneous_isotropic}
    \xi(r) = \mathbb{E} \left[ Z(\vec{x}) Z(\vec{y}) \right], \quad \text{where} \quad r = |\vec{x} - \vec{y}|.
\end{equation}

Despite being called a correlation function, what this function actually measures is something closer to the covariance. If the random field $Z(\vec{x})$ has zero mean, then Equations \ref{eq:2pcf_homogeneous} and \ref{eq:2pcf_homogeneous_isotropic} are exactly the same as the covariance function that we define in Equations \ref{eq:cov_fn} and \ref{eq:isotropic_cov_fn}. In practise, cosmologists always subtract the means from their fields before they compute the two-point correlation functions. Provided that this step is performed, the resulting two-point correlation function of the mean-subtracted random field will be exactly the covariance function of the original random field. 

We warn the reader that it is not uncommon to see $\xi(\vec{r})$ defined differently. In Equation 33.2, \citet{Peebles1980} defines the two-point correlation function (for a second-order stationary, but not necessarily isotropic random field) as:
\begin{equation}
\label{eq:2pcf_Peebles_def}
    \xi(\vec{r}) =  \frac{\mathbb{E} \left[ \left( Z(\vec{x}+\vec{r}) - \mathbb{E} \left[Z(\vec{x}+\vec{r}) \right] \right) \left( Z(\vec{x}) - \mathbb{E} \left[Z(\vec{x}) \right] \right)  \right]}{\mathbb{E} \left[Z(\vec{x}) \right]^2}.
\end{equation}
This is equivalent to the covariance function defined in Equation \ref{eq:cov_fn}, but it has been normalised by dividing by the mean of the random field squared instead of the variance. Like the correlation function (Equation \ref{eq:corr_fn}), it is unitless; but unlike the correlation function, it is not normalised to lie between $-1$ and $1$. Furthermore, this function cannot be calculated when $\mathbb{E} \left[Z(\vec{x}) \right]=0$, as we would be dividing by zero. 
\end{tcolorbox}
\newpage

\begin{tcolorbox}[colback=red!5!white,colframe=red!75!black]

Other authors \citep[e.g.][]{KT18} define the two-point correlation function $\xi(\vec{r})$ to be precisely the correlation function as we define it in Equation \ref{eq:corr_fn}. Others \citep[e.g.][]{Li+23} define it as something that is equivalent to Equation \ref{eq:corr_fn} if and only if the field in question has zero mean. In cosmology courses, it is commonly explained as the ``excess probability" $dP$ of finding a galaxy in an infintesimal volume $dV$ at a distance of $r$ from another galaxy: 

\begin{equation}
    \label{eq:prob_def_of_2pc}
    dP = n[ 1 + \xi{(r)}] dV,
\end{equation}

which is intuitively the information that is provided by all of the functions defined above. However, as a mathematical statement, this definition is not consistent with any of the definitions given above. Because the definition of this function is not universally agreed upon, we prefer to refer to the covariance function and correlation function that we explicitly define as extensions of the standard statistical concepts of correlation and covariance throughout the remainder of this Note.
\end{tcolorbox}

\newpage
\begin{tcolorbox}[colback=red!5!white,colframe=red!75!black]{
  \textbf{Wait, isn't this the autocorrelation function?} \\ 
  Sadly, the answer is that \textit{it depends who you ask}. 
  
  Lots of different definitions exist for the autocorrelation function. The one thing that almost everyone agrees on is that it is the \textit{cross-correlation} of a random field with itself. The cross-correlation of two random fields  (also known as a \textit{time series} if $\vec{x}$ is one-dimensional) $Z_1(\vec{x})$ and $Z_2(\vec{x})$ is sometimes defined to be:
  \begin{equation}
    \label{eq:xcorr_def1}
      (Z_1 \star Z_2) (\vec{r}) = \mathbb{E} \left[ Z_1(\vec{x}) Z_2(\vec{x}+\vec{r}) \right]
  \end{equation}
  In signal processing, the separation $\vec{r}$ is often referred to as the \textit{lag}. This term has also been adopted by geostatisticians. Here, the average is taken over all possible values of $\vec{x}$ for which both $Z_1(\vec{x})$ and $Z_2(\vec{x}+\vec{r})$ are known. If this definition is used, then the \textit{autocorrelation} (the cross-correlation between $Z_1(\vec{x})$ and itself) is exactly the two-point correlation function as defined in Equation \ref{eq:2pcf_homogeneous}.

  Another definition for the cross-correlation is:
 \begin{equation}
    \label{eq:xcorr_def2}
      (Z_1 \ast Z_2) (\vec{r}) = \sum Z_1(\vec{x}) Z_2(\vec{x}+\vec{r})
  \end{equation}
  If this definition is used, then the cross-correlation between $Z_1(\vec{x})$ and $Z_2(\vec{x})$ is exactly equivalent to the \textit{convolution} of $Z_1(-\vec{x})$ and $Z_2(\vec{x})$.\footnote{If $Z_1(\vec{x})$ is a complex-valued random field, then you actually need to take its \textit{complex conjugate} as well as \newline flipping it. In this case, the cross-correlation between $Z_1(\vec{x})$ and $Z_2(\vec{x})$ is equivalent to the convolution of \newline $Z_1(-\vec{x})^*$ and $Z_2(\vec{x})$, where $^*$ represents complex conjugation.}

  If we subtract the mean of each random field before taking their cross-correlation, then we get a function that is often called the \textit{cross-covariance} function:
   \begin{equation}
    \label{eq:xcov}
      K_{12}(\vec{r}) = \mathbb{E} \left\{ (Z_1(\vec{x})- \mathbb{E} \left[Z_1(\vec{x})\right] )(Z_2(\vec{x}+\vec{r})- \mathbb{E} \left[Z_2(\vec{x}+\vec{r})\right])\right\}
  \end{equation}
  }
  Letting $Z_1(\vec{x})=Z_2(\vec{x})=Z(\vec{x})$, we get the \textit{auto-covariance} function.\footnote{If things weren't bad enough already, there is \textit{also} no consensus on whether these function are supposed to  \newline be spelled with a hyphen (auto-covariance, auto-correlation) or without one (autocovariance, autocorrelation).} This function is exactly the covariance function that we estimate in Equation \ref{eq:cov_fn}. 

  Finally, people often like to take the cross-covariance function and normalise it by dividing by the standard deviation of each of the random fields. Infuriatingly, this function is also called the cross-correlation: 
  \begin{equation}
    \label{eq:xcorr_def3}
      R_{12}(\vec{r}) = \frac{\mathbb{E} \left\{ (Z_1(\vec{x})- \mathbb{E} \left[Z_1(\vec{x})\right] )(Z_2(\vec{x}+\vec{r})- \mathbb{E} \left[Z_2(\vec{x}+\vec{r})\right])\right\}}{\sqrt{\text{Var}(Z_1(\vec{x}))\text{Var}(Z_2(\vec{x}))}}.
  \end{equation}
  
If this normalisation is done, then for a second-order stationary random field, computing the cross-correlation between $Z(\vec{x})$ and itself, we get a function which is called the autocorrelation of $Z(\vec{x})$ that is exactly equivalent to the correlation function that we define in Equation \ref{eq:corr_fn}.
\end{tcolorbox}

\newpage 
\section{Statistical islands of spatial variation} \label{sec:methods}
Now that we have gathered the tools we need (Equations \ref{eq:cov_fn}-\ref{eq:isotropic_corr_fn}), we are ready to set sail and explore how these concepts are connected to the terminology used on three different ``islands of study".
We define and describe the tools used by geostatisticians (Section \ref{ssec:geostats}), cosmologists (Section \ref{ssec:cosmo}),  and fluid dynamicists  (Section \ref{ssec:turbulence}) to quantify the spatial variation that they see in their data. In each Section, we briefly recount the history of the subfield's methods, what they describe, the mathematical formalism, and how they relate to the covariance and correlation functions defined in Section \ref{sec:spatial_stats}. 
If the reader is familiar with a particular one of these subfields, it might be a good idea to start from that Section. Otherwise, these three islands can be visited in any order.

\begin{figure}[b]
    \centering
    \includegraphics[width=0.85\textwidth]{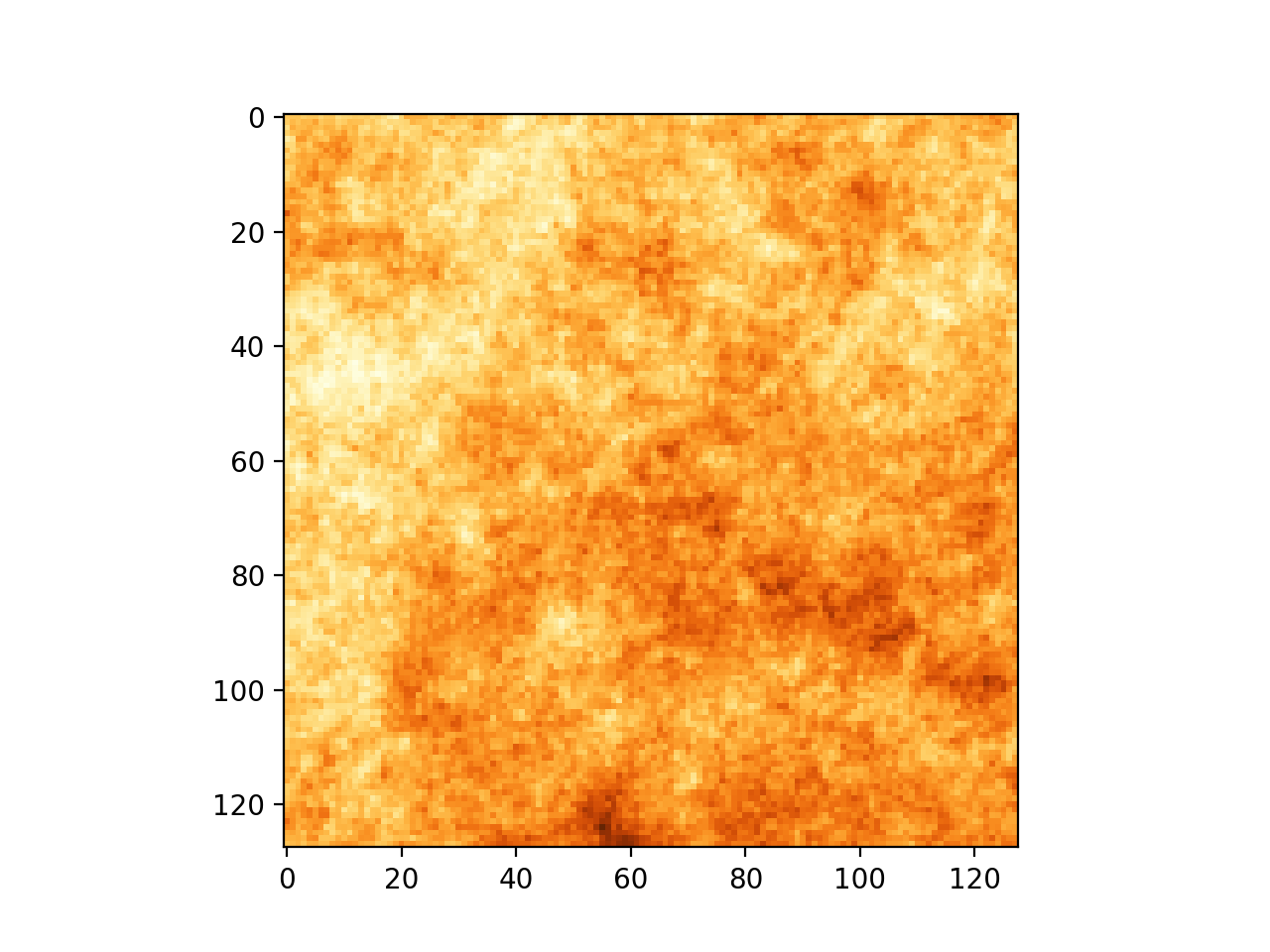}
    \caption{Introducing \textit{RaFiel}, the example \textbf{ra}ndom \textbf{fiel}d who we will be taking with us on our explorations. We will use methods native to different disciplines to analyse this field in the subsequent sections in order to determine RaFiel's second-order spatial structure.}
    \label{fig:RaFiel}
\end{figure}

Before we disembark, however, we would like to introduce our travelling companion RaFiel. RaFiel is a \textbf{ra}ndom \textbf{fiel}d. We show a picture of RaFiel in Figure \ref{fig:RaFiel}. Using Equations \ref{eq:mean} and \ref{eq:cov}, we can compute the mean of RaFiel to be $0$ and their variance to be $1$. Looking at RaFiel, they appear to be second-order stationary and isotropic, so covariance and correlation functions for RaFiel can be defined in terms of the (scalar) distance $r$ between points -- that is, Equations \ref{eq:isotropic_cov_fn} and \ref{eq:isotropic_corr_fn} can be used to fully describe their second-order structure. As we visit each island, we will describe the same random field, RaFiel, with methods native to each domain, to highlight how they are similar and different.

\subsection{The Geostatistician's Approach}
\label{ssec:geostats}

The first island we will visit is the land of \textit{geostatistics}. In the unfortunately colonial way that is all too often seen in the history of statistics, geostatistics was born out of the African mining industry in the 1950s, when a French geologist named Georges Matheron decided to take a statistical approach to prospecting \citep{matheron_history}. The original purpose of geostatistics was to give scientific answers to questions like ``Given that we see a high-grade block of ore in one location, and a low-grade block of ore in a second, where should we dig next if we want to find the best, most mineral-rich mining locations?", and ``How many core samples need to be dug out before we can understand how the gold is distributed throughout a gold field?"

Matheron found the classical statistical techniques that were used in his field to be lacking, as these approaches are not able ``to take into account the spatial aspect of the phenomenon, which is precisely its most important feature" \citep{matheron1963}. Basing his work partially on the notes of a South African mining engineer, Danie Krige, Matheron formalised the foundations of geostatistics -- the subfield of mathematics and statistics that is concerned with the analysis of random processes that vary over continuous spatial domains in a stochastic, yet predictable, way.

Since its inception, the geostatistical approach for spatial data analysis has been put to use in a lot of diverse fields, including epidemiology, climate modelling, ecology, economics, and of course geology. The use of geostatistical methodologies applied to astronomical data is an active area of study \citep{Clark+19, Gonzalez-Gaitan+19, Geogals1}.
We include the semivariogram in this Note for two reasons. Firstly, it requires less mathematical complexity to construct than the subsequent methods. This makes it easier to compare to the fundamental statistics we describe in Section \ref{sec:classical_stats}. The second reason is that geostatistical methods show great promise in comparing quantitative theories about the Universe to observational data \citep{Geogals1}, estimating values in incomplete data sets \citep{Geogals2}, and 
may be helpful for other astronomical applications, especially with high-resolution data \citep{GeoGals3}. In this investigation, we will review one important tool from the geostatistical approach, the \textit{semivariogram}, and see what it can tell us about the correlation structure of our friend RaFiel.

\subsubsection{The Semivariogram}

The \textit{semivariogram} is a classical tool from geostatistics that is used for exploratory data analysis. It is a data visualisation method, whose purpose is to reveal the spatially correlated nature of the observed data. Intuitively, it shows how the variance between data points depends on their separation. Examination of a semivariogram plot serves as a test to check if there is spatially (or temporally) correlated structure in an observation of a random field -- and, if it exists, further examination reveals the amount of variance in the data explainable by spatially correlated effects, and estimates the spatial scales over which such correlations are effective.

To compute the semivariogram for a second-order stationary (or homogeneous, or translationally-invariant) random field, take all possible pairs of data points, $Z(\vec{x})$ and $Z(\vec{y})$, and group them by their spatial separation $\vec{r}$. We then look at the variance of the difference between each pair of values that are separated by about $\vec{r}$. Formally, it is defined as:
\begin{equation}
    \label{eq:svg}
    \gamma(\vec{r}) = \frac12 \widehat{\rm Var} \left( Z(\vec{x} + \vec{r}) - Z(\vec{x}) \right).
\end{equation}

If your random field also happens to be isotropic, then the semivariogram only depends on the distance $r = | \vec x - \vec y|$ between data points. In this case, we can simplify our expression for the semivariogram to:
\begin{equation}
    \label{eq:svg_isotropic}
    \gamma(r) = \frac12 \widehat{\rm Var} \left( Z(\vec{x}) - Z(\vec{y}) \right),
\end{equation}
where the variance is computed over all pairs of points $\vec x$ and $\vec y$ for which $r - \delta/2  \leq | \vec x - \vec y| \leq r + \delta/2$.\footnote{This function, defined over bins of separations $r$, is sometimes referred to as the \textit{empirical semivariogram} to distinguish it from the \textit{theoretical semivariogram}, which uses the exact variance (not an estimator), has no binning, and is impossible to compute from observational data. Since we're in the business of making sense of the world around us and not of defining abstract mathematical functions, we will hereafter use the term \textit{semivariogram} to mean the empirical semivariogram. Both the empirical and theoretical semivariograms are also, confusingly, often referred to as \textit{variograms} -- we warned you that the sea of definitions was treacherous!} Here, $\delta$ is the \textit{bin width} of the semivariogram. It is a \textit{hyperparameter}, meaning that there is no perfect mathematical way to select it. Really, any value will do, but you need to keep two things in mind: firstly, the semivariogram won't be able to tell you about anything that is happening on scales smaller than $\delta$ -- so if you are interested in variation on scales of tens of parsecs, choosing $\delta=100$ parsec is a bad idea. This is a reason to make your bin size smaller. Secondly, you will need a few pairs of data points in order to compute the variance in each bin reliably. This is a reason to make your bin size larger. As long as your bin size is small enough that you can see what you're interested in, and large enough that you can be pretty confident in your variance estimates at each spatial separation (common wisdom amongst statisticians states that we desire at least 30 (and preferably $\sim 50$) data point pairs at each separation; \citealt{Schabenberger+Gotway}), then you're good.\footnote{But how can I measure how confident I ought to be about my variance estimates at each location? Well, to do that, you need to estimate the uncertainty on your variance estimates at each location. That is, you need to find the \textit{variance} of your \textit{variance estimate}. Formally, this can be done - and there's nothing wrong with doing this - but people will make \href{https://xkcd.com/2110/}{jokes about you on the internet} for it, the average elevation of eyebrows in your local area will rise by several millimetres, and nobody on the \href{https://stats.stackexchange.com/questions/541888/variance-in-variance-weighted-variance-estimate}{statistics stack exchange} will take your questions seriously. Most people don't bother with this, so don't worry about it.}

\begin{figure}
    \centering
    \includegraphics[width=0.9\textwidth]{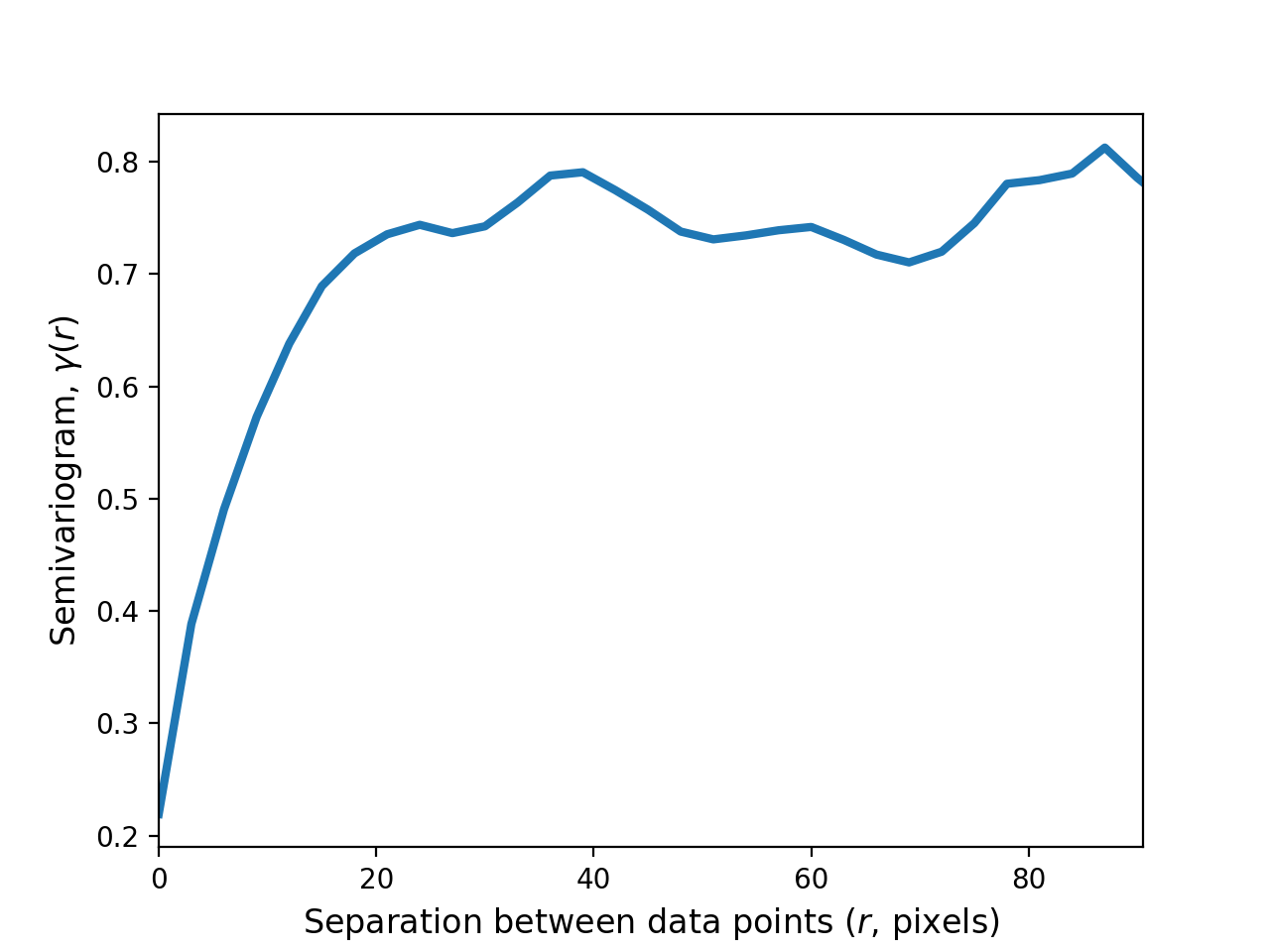}
    \caption{This is a semivariogram for RaFiel, our random field. The semivariogram increases for points that are separated by larger distances, because points that are closer to each other tend to have more similar values. This behaviour reflects what we see in the image of RaFiel (Figure \ref{fig:RaFiel}). At a separation of $r\sim 20$ pixels, the semivariogram tends to reach a near-constant value. This tells us at what distance points stop being correlated with each other (called the \textit{range} in the geostatistical literature).}
    \label{fig:svg}
\end{figure}

The best way to learn about what a semivariogram can tell us is to look at one. We show a semivariogram of RaFiel in Figure \ref{fig:svg}. Because RaFiel is isotropic, we can use the one-dimensional form of the semivariogram that we define in Equation \ref{eq:svg_isotropic}.\footnote{Be sure to check out the \href{https://github.com/sabrinastronomy/rosetta-stone}{Jupyter notebook} to see how this is done in practise. Let the soft hum of your laptop's computer fan be the peaceful white noise that soothes you as you read through the rest of this Section.}
From this Figure, we can see that the semivariogram, $\gamma(r)$, tends to increase as $r$ increases. This tells us that RaFiel obeys Tobler's First Law of Geography. For data points that are closer to each other (lower $r$), there is less variance in the difference between their values -- in other words, points that are close to each other tend to have more similar values than points that are farther away.

\newpage
At large separations ($r \gtrsim 40$ pixels), this semivariogram seems to flick up towards higher values before becoming ```wobbly". This effect is commonly seen in semivariograms. As we go to further separations, the amount of pairs available for each variance estimate decreases, and so the variability in the semivariogram increases.
It is nothing to worry about -- all of the information that we are interested in happens on spatial scales much smaller than the size of our box, before the semivariogram becomes unreliable. Unfortunately, there is no exact formula for at which separation these edge effects kick in -- but a common recommendation is to only compute the semivariogram for separations of up to $\frac12$ of the maximum separation in the data \citep{Schabenberger+Gotway}. On the other end of things, the smallest variations that a semivariogram analysis can pick up in theory are variations of sizes equal to the minimum separation between observations -- i.e. the size of an individual pixel. In practise, the size of a fluctuation must be greater than about twice the size of a pixel to be accurately captured using a geostatistical analysis framework \citep{GeoGals3}.
\newpage
By looking more closely at the semivariogram, we can read three important parameters that give us a broad overview of the way that this data field is spatially correlated: 
\begin{itemize}
    \item The \textit{range} is the separation between data points at which the variance appears to stop increasing. There is no precise mathematical way that this is calculated -- instead, we just look at the data, and see where, approximately, it starts to get flat. This tells us where the largest spatial variations in the data can be found. At spatial locations separated by more than the range, for practical purposes, data points will be uncorrelated.
    \item The \textit{sill} is the maximum height that the semivariogram reaches, or the height that it flattens out to at its range. This value is equal the total variance in the data, as would be computed from Equation \ref{eq:var}.
    \item The \textit{nugget} is the inferred height of the semivariogram as $r \to 0$. This quaint term gets its origins from the gold fields where geostatistics was first hit upon -- the size of a nugget of gold was much smaller than the size of the rock samples that were analysed, so the presence or absence of these nuggets would cause large variations in the amount of gold found between samples, no matter how close they were to each other \citep{chiles_delfiner99}. This is an 
    example of a \textit{microscale variation} -- a variation that happens on scales smaller than the size of a single sample of the data. We cannot see this variation with a geostatistical approach, as we cannot make out any variations smaller than a single pixel.
\end{itemize}

In addition to being a handy tool for visualising the spatial structure of a random data field, the semivariogram also satisfies several nice mathematical properties. For second-order stationary random fields (that is, random fields for which Condition \ref{eq:weak-sense-stationary} is true for all pairs of points), the following relationship holds:
\begin{equation}
    \label{eq:svg_to_cov}
    \gamma(\vec{r}) = C(0) - C(\vec{r}),
\end{equation}
where $C(\vec{r})$ is the covariance function defined in Equation \ref{eq:cov_fn}. 
If we want to, we can manipulate this expression to relate the semivariogram to the correlation function, $\rho(\vec{r})$, and the total variance, $\sigma^2$, of the data:
\begin{equation}
    \label{eq:svg_to_2pc}
    \gamma(\vec{r}) = \sigma^2\left(1-\rho(\vec{r})\right).
\end{equation}
So the semivariogram is just a manipulated version of the covariance function. It is related to the correlation function, but captures one extra piece of information: the total variance present within the random field. We will see how this function relates to other approaches used in astronomy in subsequent Sections.

\subsection{The Cosmologist's Approach}
\label{ssec:cosmo}

The Fourier approach to quantifying spatial variations within cosmology and large scale structure studies arose through the need to learn about structure on large scales with minimal information. Modern cosmology began with Albert Einstein's general theory of relativity \citep{Einstein1915} and Edwin Hubble's discovery of the expanding Universe \citep{Hubble1929}. Both of these theories make predictions about the relationship between and evolution of structures in our Universe on various distance scales. 
To truly understand our universe on all scales, we need to study how often different scales appear within our universe -- that is, we need to know how \textit{correlated} structures in our Universe are.

In the mid 1900s, cosmologists discovered that the distribution of galaxies on the sky was not uniform. Instead, galaxies were seen to \textit{cluster}. If one patch of the Universe was seen to be rich in galaxies, then nearby patches were found to be more likely to also be rich in galaxies \citep{1934ApJ....79....8H, 1934Natur.133..578B, 1938PASP...50..275M}. In other words, on intergalactic scales, the Universe was seen to obey Tobler's First Law of Geography.

Zooming out even further, George \citet{1958ApJS....3..211A} predicted that galaxy clusters themselves exist within \textit{superclusters} -- clusters of galaxy clusters. This was accomplished by assuming all clusters were distributed randomly with no correlation and calculating the probability (using classical statistical methods) of observing the population counts of galaxies that he found in his data. Abell found that the probability of the data having no spatial structure was virtually impossible, but still had no way to quantify what the actual spatial structure was.\footnote{This kind of calculation is called a \textit{p-value}, where you assume that something you think is wrong (a \textit{null hypothesis}) is true, and then calculate the chances of it being right. If you find that there's a low probability of the thing you think is wrong being right (a low p-value), then you can be pretty sure that the thing you think is wrong really is wrong (you can \textit{reject the null hypothesis}). Trouble is, this approach won't tell you what the right thing to believe instead is.}

Disappointed by the lack of good spatial statistical methods in their field
, two cosmologists, graduate student Jer Tsang Yu and his supervisor James Peebles, came up with a clever way to quantify information on various spatial scales \citep{1969ApJ...158..103Y}.
Drawing inspiration from the concept of \textit{power spectral density} that was recently developed for signal processing applications \citep{blackman_tukey}, Yu and Peebles made the first application to quantify clustering length scales with the \textit{power spectrum}.

There now exists no cosmology textbook in the Universe that fails to mention the power spectrum both as a function of spatial modes (or Fourier modes, $k$, defined in Section \ref{subsec:kmode_pspec}) and angular modes (or multipole moments, $l$)\footnote{Often, cosmologists refer to the power spectrum as a two-point statistic, meaning it uses two points at a time to quantify correlations. The trispectrum (See Glossary) is an example of a higher order statistic.}. The latter enables us to study the cosmic microwave background (CMB)\footnote{In 1965, two radio astronomers working at Bell telephone laboratories, Arno Penzias and Robert Wilson, were vexed by a mysterious, constant microwave signal in their data. After evicting a family of pigeons that were living in their satellite dish and cleaning out their droppings, the signal remained. Fortunately, Penzias and Wilson were in contact with physicists from MIT who could identify what this signal really was: a signature of leftover radiation from moments after the Big Bang, when the Universe was incredibly hot and dense. Dubbed the \textit{cosmic microwave background} (or CMB for short), this signal tells us a lot of information about what the Universe was like only 380,000 years after the Big Bang \citep{1965ApJ...142..419P}.}, while the former most commonly allows for study of the evolution of large scale structures in our Universe. Both of these tools have been used to develop insight into the evolution of our Universe in synergistic ways. 
\newpage
\subsubsection{The k-mode Power Spectrum}
\label{subsec:kmode_pspec}

In order to define what a power spectrum is, we must first introduce the Fourier transform -- a way to think about signals as the sums of many different, independent waves. Fourier transforms were invented by the French mathematician and physicist Jean-Baptiste Joseph Fourier.\footnote{J. J. Fourier was a bit of a character. In addition to inventing the Fourier transform, he was also a governor of Lower Egypt in Napoleon's army, is credited with discovering the greenhouse effect, and enjoyed wrapping himself in a warm blanket and walking around his mansion in it. This last hobby led to his tragic death in 1830, when, wrapped in a blanket, he fell down his stairs and was unable to break his fall \citep{quantum_universe}. At least he died doing what he loved.} In a nutshell, Fourier's theorem is this: any signal, no matter how complicated the shape as long as its reasonably well-behaved\footnote{Determining whether a Fourier transform exists for a signal is actually quite complex, and depends on the integrability of the function. We direct the reader to \citet{Champeney_1987} for a more rigorous explanation of the conditions required for a Fourier transformation to be performed.}, can be expressed as the sum of many different \textit{waves}. The \textit{Fourier transform}, $\mathcal{F}$, takes a signal that is a function of $\vec{r}$ as an input, and figures out what combination of waves are needed to reproduce that signal. We describe these waves in terms of their \textit{wavenumber}, $\vec{k}$. A wave with a wavenumber of $\vec{k}$ oscillates along the direction of $\vec{k}$, and will undergo one full cycle over a distance of $\frac{2\pi}{|\vec{k}|}$. The term \textit{k-mode} is used to refer to a wave with a wavenumber of magnitude $k$.

The Fourier transform of a function $f(\vec{r})$ is written with a squiggly hat, $\Tilde{f}(\vec{k})$. The value of $\Tilde{f}$ at each wavenumber $\vec{k}$ tells you how much of that particular wave needs to be added in order to reproduce your spatial signal, $f(\vec{r})$. We say that the function $\Tilde{f}(\vec{k})$ exists in \textit{Fourier space}, or \textit{k-space}, and the function $f(\vec{r})$ exists in \textit{real space}, or \textit{configuration space} -- but really, they are the same function. Both $f(\vec{r})$ and $\Tilde{f}(\vec{k})$ contain the same information. This is essentially the same as the two functions being expressed in two different bases.

To make these concepts more concrete, it helps to have an example that we are familiar with. In classical music, orchestras all tune up by playing the same note: the A above middle C. To make sure everyone is always playing the same note, this pitch is defined to have a frequency of exactly 440Hz.\footnote{You can hear this pure, $440$Hz, concert pitch A tone \href{https://upload.wikimedia.org/wikipedia/commons/5/50/Sine_wave_440.ogg}{here}.} Taking a Fourier transform of this signal, we would see that $\Tilde{f}(\vec{k})$ has a large spike at a value of $\vec{k}=+440$ Hz, and is zero everywhere else -- this signal can be made up of one wave alone. If we were instead to play this concert A on a piano and feed the time series of the sound into a spectrometer (a machine that lets us visualise amplitude versus frequency as a function of time), we would see some amount of power at the \textit{harmonic} frequencies of integer multiples of $440$ Hz (at $880$ Hz, $1320$ Hz, $1760$ Hz, and so on). If we repeated this experiment with a flute, or a trumpet, or a cello, we would see that each of these different harmonics contributes a different amount to the final signal. By looking at the spectral signatures revealed by taking the Fourier transform of our signals, we could clearly distinguish between which instrument is being played.

To put this concept into precise mathematical language, we define the Fourier transform, $\mathcal{F}$, for a 3-dimensional function $f(\vec{r})$ as follows:\footnote{We stick with the cosmologist's way of expressing the Fourier transform, but other fields use different conventions. For example, you might see the negative sign in the exponent of the inverse Fourier transform or different placements of $2\pi$. In this \href{https://www.wtamu.edu/~dcraig/PHYS4340/070413_FTconventions.pdf}{PDF} with common conventions, we use the ``standard layout".}

\begin{equation}
    \label{eq:fft}
    \Tilde{f}(\vec{k}) = \mathcal{F}\{f(\vec{r})\} = \int_{-\infty}^{\infty} e^{-i\vec{k} \cdot \vec{r}} f(\vec{r})dr^{3} .
\end{equation}
The Fourier transform is an information-preserving operation -- that is to say, both $f(\vec{r})$ and $\Tilde{f}(\vec{k})$ tell us everything that there is to know about a signal. Because of this, we can also use the Fourier transform of a signal, $\Tilde{f}(\vec{k})$, to figure out what the original signal was. This is done through an operation called the \textit{inverse Fourier transform}, which can be calculated with the following formula:
\begin{equation}
    \label{eq:ifft}
    f(\vec{r}) = \mathcal{F}^{-1}\{\Tilde{f}(\vec{k})\} =  \int_{-\infty}^{\infty} \frac{d^{3} k}{(2\pi)^3} e^{i\vec{k} \cdot \vec{r}} \Tilde{f}(\vec{k}).
\end{equation}
Interestingly, the formula for taking the inverse Fourier transform of a function in $k$-space (Equation \ref{eq:ifft}) looks very similar to the formula that we use to take a Fourier transform of a function in real space (Equation \ref{eq:fft}). With the definition of the Fourier transform commonly used in cosmology, the key things to be careful about are the extra factors of $2 \pi$ in the denominator of the inverse Fourier transform, and the change to a positive sign in the exponent going from Equation \ref{eq:fft} to \ref{eq:ifft}. To probe the spatial analog to musical frequency in Fourier space, we use $k$-modes which correspond to a length scale of $2\pi/|\vec{k}|$ Essentially, the Fourier transform takes our data in terms of its actual value at each position and converts it into an amplitude of different scales in our data.

We define the power spectrum of a second-order stationary (or homogeneous, or translationally-invariant) random field as the Fourier transform of the covariance function of that field:\footnote{If you read any cosmology textbook on the planet, you will probably see the power spectrum instead defined as the Fourier transform of the \textit{two-point correlation function}, $\xi(\vec{r})$. \textbf{This is exactly equivalent to what we are doing here.} We choose to write it in this way for two reasons. Firstly, the two-point correlation function actually measures \textit{covariance} for a zero-mean field, not correlation as defined in Equation \ref{eq:corr}. Secondly, nobody outside of cosmology knows what a two-point correlation function is. If you are a cosmologist who wants to be able to explain your methodology to people outside of astronomy, this is the least confusing way to do it that we could think of.}

\begin{equation}
    \label{eq_matter_pspec_full}
    P(\vec{k}) \equiv \int^{\infty}_{-\infty} dr^{n} e^{-i  \vec{k} \cdot \vec{r}} C(\vec{r}),
\end{equation} where n is the number of dimensions of our random field. For an isotropic (statistically symmetric) 1D random field (i.e. time series data), this equation simplifies to:

 \begin{equation}
    \label{eq_1d_matter_pspec}
    P(k) =  2 \int^{\infty}_{0}  C(r) \cos ( kr) dr.
\end{equation} 
In this case, the resulting power spectrum is often called the \textit{power spectral density} (or some permutation of those three words -- see the Glossary).

For an isotropic, 2D field (such as RaFiel), then we can obtain the power spectrum from an integral over one spatial dimension as well:

 \begin{equation}
    \label{eq_2d_matter_pspec}
    P(k) = 2\pi \int^{\infty}_{0}  C(r) J_0(kr) rdr ,
\end{equation} 
where $J_0 (x)$ is the zeroth-order Bessel function of the first kind. If we instead consider a 3D field that is also isotropic, then the power spectrum simplifies to:
 \begin{equation}
    \label{eq_matter_pspec}
    P(k) = 4\pi \int^{\infty}_{0}  C(r) \frac{\sin (kr)}{kr} r^2dr.
\end{equation} 
We skip the derivation, but you can find it \href{https://people.ast.cam.ac.uk/~pettini/Intro%20Cosmology/Lecture14.pdf}{here}. Equation \ref{eq_matter_pspec} is the $k$-mode power spectrum for a second order stationary, isotropic field. We're still just one Fourier transform away from the covariance function we defined in Equation \ref{eq:cov_fn} in Section \ref{sec:classical_stats}. Because the Fourier transform is an invertible transform, the power spectrum contains exactly the same information as the covariance function does -- if the power spectrum of a random field is known, then its covariance function can be reconstructed. We can go one step further, and convert the Fourier transform of the power spectrum to the semivariogram using Equation \ref{eq:svg_to_cov}. Since this equation is also invertible, these two equations form a bridge that allows the methodology used to capture second-order structure by cosmologists to be translated into the methodology used by geostatisticians: both approaches capture the exact same information about the random fields under investigation. 

In the ideal case, when we move our quantification of spatial correlations into Fourier space, we are only capturing variances of $k$-modes, i.e., we are diagonalising the covariance matrix. When we look at the evolution of the matter distribution in our Universe to first order through the lens of the power spectrum, only the amplitude of the power spectrum changes as a function of time. The $k$-modes in the matter power spectrum of our Universe evolve independently. This convenience exemplifies the utility of the power spectrum within cosmology. 
\begin{tcolorbox}[colback=red!5!white,colframe=red!75!black]
  \textbf{Wait, how does the power spectrum diagonalise the covariance matrix in Fourier space?} \\
We know that the power spectrum is the Fourier transform of the covariance function of our data (or the two-point correlation function, or the autocorrelation function, \textit{depending on who you talk to} -- see the red \textbf{Wait!} boxes in Section \ref{sec:spatial_stats}), but it \emph{also} presents interesting consequences in Fourier space by effectively removing covariance in k-space.
We can make this connection explicit with the following Equation, which holds true for second-order stationary random fields ($Z(\vec{x})$ for which Condition \ref{eq:weak-sense-stationary} holds true for all pairs of points $\vec{x}$ and $\vec{y}$):
    \begin{equation}
    \label{eq:iso_pspec}
    \text{Cov} (\Tilde{Z}(\vec{k} + \vec{\Delta k}), \Tilde{Z}(\vec{k})^{*}) = (2\pi)^D \delta(\Delta \vec{k}) P(\vec{k}),
    \end{equation}
where $\delta(\vec{\Delta k})$ is the Dirac delta function, which is equal to 0 everywhere except when the difference between $k$-modes $\vec{\Delta k} = 0$, and $D$ is the dimensionality of our data.
This shows that when we move to Fourier space, the $k$-space representation of its covariance matrix is diagonal. Any off diagonal elements, i.e. elements for which the difference between $k$-modes $\vec{\Delta k} \ne 0$, will have covariances of zero in Fourier space. That is, there will be \textbf{no correlation} between different $k$-modes in Fourier space. Equation (\ref{eq:iso_pspec}) comes from the fact that our data is translationally invariant, and the nature of the Fourier transform itself. If the random field $Z(\vec{x})$ was not statistically translationally invariant to second-order, then there would still be correlations between different $k$-modes in Fourier space. For further details and a proof of this, see Section 4 of this \href{https://github.com/acliu/21cmStarterKit/blob/57e1af5ca9c0b8b098ca35942278439daaf94a72/Basics/PowerSpectra.pdf}{Note}. Also, in real observations, we're never actually perfectly diagonalising covariance matrices due to the effects of foregrounds, galactic extinction, survey geometry, and instrumental effects, which can be collectively encapsulated in functions termed \textit{window functions} \citep{Liu_2011, Karim+23}.
\end{tcolorbox}

The computational implementation and mathematical formalism for calculating a power spectrum may not seem equivalent on first glance, but Equation \ref{eq:iso_pspec} provides the first glimpse into its numerical calculation. Constructing the 3D power spectrum numerically requires summing up power in spherical shells in Fourier space. We show how this is implemented computationally in \href{https://github.com/sabrinastronomy/rosetta-stone/blob/main/rosetta_stone.ipynb}{our Jupyter notebook tutorial}. In addition to the convenience of describing the matter evolution of our Universe through the matter power spectrum, Fast Fourier Transforms (FFTs) allow us to compute the Fourier transform of our fields and thus the power spectrum extremely efficiently. 

In a real world computation of the power spectrum, there are limits to the minimum and maximum spatial scales that we can probe. Consider the Fourier transformed version of our field in $k$-space. Although we could in theory calculate infinitesimally small $k$-modes, the smallest $k$-mode that we can learn about corresponds to the size of the field of view of our data, $L_{\rm max}$ -- that is, $k_{\rm min} = \frac{2 \pi}{L_{\rm max}}$.
At this spatial frequency (i.e. $k$-mode), each oscillation covers exactly one pixel in Fourier space. Decreasing the frequency of these waves would create multiple oscillations over the same Fourier space pixel and does not provide any new information. On the (real space) small scale or (Fourier space) larger $k$-mode end, we could also theoretically take larger and larger $k$-modes. When computing the power spectrum, our maximum useful $k$-mode is $\frac{2 \pi}{\sqrt{n}~\rm pixel~size}$, where $n$ is the number of dimensions of our field. 
For example in 2D, the largest wave we can form in our square box in Fourier space will have a magnitude of $\textbf{k} = \sqrt{k_x^2 + k_y^2}$, which is $\textbf{k} = \sqrt{2k^2}$ because of isotropy. In the actual computation of a 2D power spectrum, this is the same as putting an upper limit on the size of the circle in Fourier space in the circular binning that we use.

\begin{figure*}
    \centering
    \includegraphics[width=0.8\textwidth]{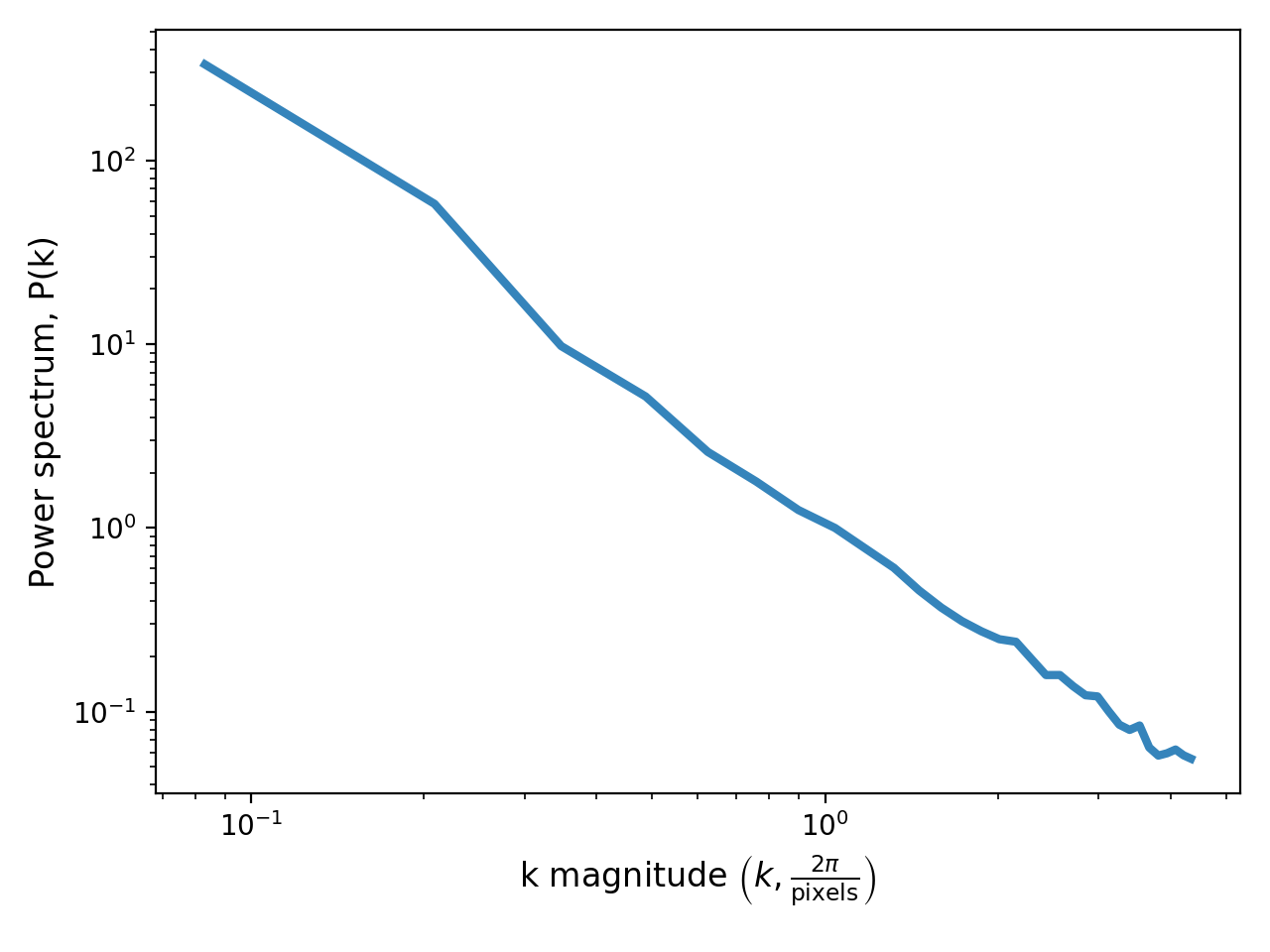}
    \caption{This is the power spectrum for RaFiel, our random field. We express the $k$ magnitudes following the cosmological Fourier transform conventions. The $k$ magnitude units are $2\pi/\rm pixels$. Most of our power is at small k-modes (large spatial scales). This makes sense as we can see significant large scale structure in RaFiel extending over many pixels (Figure \ref{fig:RaFiel}). The largest $k$-mode that we show represents the smallest scale that you can probe in your field, which is $\sqrt{2}$ pixels (Section \ref{subsec:kmode_pspec}). We can only see correlations that are larger than that size.}
    \label{fig:pspec}
\end{figure*}

A power spectrum of RaFiel can be seen in Figure \ref{fig:pspec} where we show the power spectrum as a function of $k$-magnitude $(\frac{2 \pi}{\rm pixels})$. After learning about semivariograms, the x-axis of Figure \ref{fig:pspec} will be backwards to your intuition: the largest spatial scales correspond to the $k$-modes closest to zero, and the smallest spatial scales are the largest k-modes. We see more noise in the power spectrum at the largest $k$-modes, as it gets harder and harder to probe smaller scales as we reach down towards our resolution limit.

\subsection{The Fluid Dynamicist's Approach}
\label{ssec:turbulence}

The third island in our whirlwind tour of spatial data analysis methodologies is the world of fluid dynamics, where random fields are usually a result of \textit{turbulence}. 

Turbulent flows are spatially and temporally stochastic. They are uneven, unstable and unpredictable, with large, irregular variations in the fluid velocity appearing over a wide range of spatial scales. Despite these difficulties, understanding turbulence is important in science and engineering for a lot of reasons -- for example, to make accurate predictions of the weather, to understand the mechanisms that quench star formation in the interstellar media of galaxies, and to make sure that our aeroplanes don't fall out of the sky.

While the equations that govern turbulent fluids are well-known and easy to write down, trying to solve them (or even knowing if they can be solved) is a million dollar question.\footnote{We're not kidding. At the turn of our millennium, the Clay Mathematics Institute of Colorado put a million dollar bounty on \href{https://www.claymath.org/millennium-problems/}{seven mathematical problems} that they wanted to see solved in the next thousand years. Finding out if smooth solutions always exist to the Navier-Stokes equation, which describes how turbulent flows evolve with time, is one of these problems.} To make progress on this problem, two steps had to be made. The first was to consider the \textit{statistical properties} of the pressure, velocity, and temperature of the resulting turbulent field, rather than trying to model how they evolve explicitly. This is what we have been doing with all of the random fields that we have encountered so far already, so this step is not new to us. The second step was made by British meteorologist Lewis Fry Richardson\footnote{In addition to being a meteorologist, Richardson was a pacifist, in a very mathematical way. After developing the mathematical technology that is used to predict the weather, Richardson attempted to use that same maths to model how wars start, in order to figure out how to prevent them. While he was not completely successful in preventing all future wars, he did end up writing a very interesting book on the topic \citep{Richardson60}.}, who (to the best of our knowledge) came up with the theory of the turbulent \textit{energy cascade} \citep{Richardson1922}. In this theory, turbulence begins with large-scale eddies, which are unstable, and break up into smaller eddies, which are unstable, and break up into smaller eddies on smaller scales still. This process stops when the eddies become so small that the random motion of particles becomes comparable to the sizes of the eddies, at which point the turbulent kinetic energy diffuses into thermal energy. This theory is summarised in Richardson's famous little poem: \textit{Big whirls have little whirls that feed on their velocity, and little whirls have lesser whirls, and so on to viscosity} \citep{Richardson1922}.

The key takeaway from Richardson's theory is this: the scale over which these turbulent fluctuations occur is important. This paved the way for two Soviet mathematicians, Andrey Kolmogorov and his collaborator Alexander Obukhov, to explore how the variation in the velocity of a turbulent medium changes over different spatial scales \citep{Yaglom90}. In 1941, both of these mathematicians delivered groundbreaking results on how turbulence is expected to behave in a statistical sense over a range of spatial scales. However, in a historical twist that will shock absolutely no one by this point, both mathematicians chose to present their results using different tools and different terminology. \citet{Obukhov41} presented his results on the way that turbulence is structured in Fourier space, using a tool called the \textit{energy spectrum} that is \textbf{exactly the same as the power spectrum} that was explained in Section \ref{subsec:kmode_pspec}. On the other hand, \citet{Kolmogorov41} stayed in real space, and computed the spatial correlation structure of turbulent velocities using a tool that is now known as the \textit{structure function}. To this day, both structure functions and energy spectra are used in turbulence analysis.

\subsubsection{The Structure Functions}

The $p$-th order structure function of a scalar-valued random field $Z(\vec{x})$ at a separation of $\vec{r}$ is defined to be the average absolute difference between values of the random field at points separated by $\vec{r}$ raised to the power of $p$, where $p$ is a positive integer:
\begin{equation}
    \label{eq:scalar_structure_func_p}
    S_p(\vec{r}) = \mathbb{E}\left[ { \left( | Z(\vec{x} + \vec{r}) - Z(\vec{x}) | \right)}^p \right].
\end{equation}
In other words, it is the $p$-th moment of the absolute  difference between $Z(\vec{x} + \vec{r})$ and $Z(\vec{x})$.\footnote{There are other ways to compute these structure functions -- for example, rather than using pairs of points at two locations ($\vec{x}+\vec{r}$ and $\vec{x}$), you could also use a triple of three points ($\vec{x}+\vec{r}, \vec{x},$ and $\vec{x}-\vec{r}$) to compute the structure functions of any order. For details on why you would want to do it this way and equations for how this is done, see \citet{Seta+23}.} In the case where $Z(\vec{x})$ is isotropic, then this value will only depend on the magnitude of $\vec{r}$ and not its direction, and so we can define the structure function as:
\begin{equation}
    \label{eq:scalar_structure_func_isotropic}
    S_p(r) = \mathbb{E}\left[ {\left(| Z(\vec{x}) - Z(\vec{y}) | \right)}^p \right], \, \text{where} \, |\vec{x}-\vec{y}|= r.
\end{equation}

The third, fourth, and higher order structure functions tell you about the third, fourth, and higher moments of your random field -- but since we are only focusing on second-order spatial structures of random fields in this Note, we limit our discussion to the structure functions of the first and second order and its comparison to other two-point statistics: the semivariogram (Section \ref{ssec:geostats}) and the power spectrum (Section \ref{ssec:cosmo}).

The first-order structure function seeks to achieve the same purpose as the semivariogram and power spectrum -- to quantify the spatially-correlated nature of variability within the data as a function of scale. However, unlike the semivariogram, which looks for the variance between pixels at a given separation, the first-order structure function is based around a different measure of spread -- namely, the \textit{mean absolute deviation} between pairs of values. In spirit, this measurement is similar to a standard deviation up to a small correction factor -- for Gaussian distributions, the size of the mean absolute deviation between pairs is $\frac{2}{\sqrt{\pi}} \sigma$, or approximately $1.13\sigma$. However, the first-order structure function is not commonly used to quantify spatial correlations, as its results cannot be readily converted into statements about variances, standard deviations, or correlations (i.e. the useful statistics described in Section \ref{sec:spatial_stats}) for random fields in general.

The second-order structure function instead tells us about how the covariance between data points $Z(\vec{x} + \vec{r})$ and $Z(\vec{x})$ depends on their separation, $\vec{r}$. If this sounds similar to a semivariogram, that's because it is: for second-order stationary data or data with zero mean, the second-order structure function is related to the semivariogram by the following equation:
\begin{equation}
    \label{eq:svg_to_structure_function}
    S_2(\vec{r}) = 2 \gamma(\vec{r}).
\end{equation}
 Combining this insight with Equation \ref{eq:svg_to_cov}, we can relate the second-order structure function to the covariance function which is defined for second-order stationary random fields in Equation \ref{eq:cov_fn}:
\begin{equation}
    \label{eq:structure_function_to_cov_fn}
    S_2(\vec{r}) = 2 \left(C(0) - C(\vec{r})\right).
\end{equation}
Using this, we can in turn express the second-order structure function in terms of the correlation function, $\rho(\vec{r})$, that we define for second-order stationary random fields in Equation \ref{eq:corr_fn} and the total variance $\sigma^2$:
\begin{equation}
    \label{eq:structure_function_to_corr_fn}
    S_2(\vec{r}) = 2 \sigma^2 \left(1 - \rho(\vec{r})\right).
\end{equation}
So the second-order structure function is \textit{also} just a manipulated version of the covariance function. This shows us that the second-order structure function gives \emph{exactly} the same information that a semivariogram does! Because we already showed how power spectra and semivariograms contain the same information, this means that the second-order structure function also gives the exact same information that a power spectrum (or an \textit{energy spectrum} as it's called in the turbulence literature) does.

\newpage
\section{Connecting the islands} 
\label{sec:reflections}

\begin{figure}
    \centering
    \includegraphics[width=\textwidth]{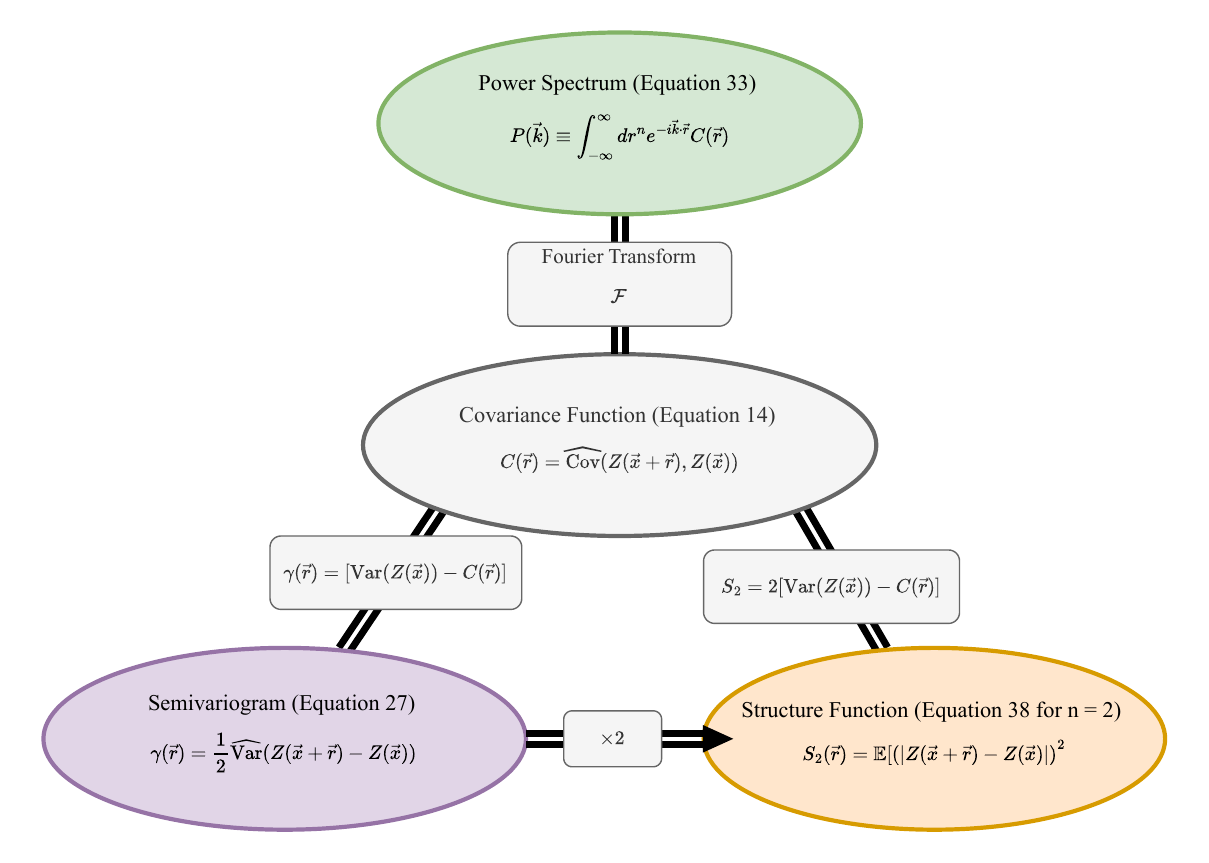}
    \caption{This is the Rosetta Stone that we have been searching long and hard for. We summarise how all methods described in this Note are directly related to the covariance function, which shows how they can all be translated into each other. Note that all of these formulae assume a second-order stationary, real valued random field.}
    \label{fig:methods_flow}
\end{figure}

All of the three methods that we have explored (\textit{power spectra}, \textit{semivariograms} and \textit{second-order structure functions}) give the same information. All of them can be transformed into the \textit{covariance function}, which is the same as the \textit{two-point correlation function} for a zero-mean field. Dividing by the variance, we get the \textit{correlation function}, which is sometimes (but not always) the same as the \textit{autocorrelation function}, depending on who you ask. We show how all of these different functions are connected in Figure \ref{fig:methods_flow}.

A natural question to ask after having read this note is, ``When should I use each of these methods?" Unless you're applying a new statistical technique, you will probably end up applying whichever one is already used to your subfield. Nonetheless, we provide a brief overview of when to use the power spectrum versus another method described in this Note not requiring the Fourier transform. 


To avoid Fourier space, the semivariogram or the structure function present themselves as the ideal alternative to the power spectrum. The benefits of these methods is that both deal very well with random fields that have many missing data points, which are commonly encountered in astronomical problems. These methods also benefit from their explainability, as they require less mathematical underpinning than the Fourier methods.

Conversely, the methods that employ the Fourier transform (the energy spectrum or power spectrum) benefit from speed and computational efficiency. For a semivariogram (or second-order structure function) to be computed, a distance matrix must be constructed between all pairs of data points, which can use up a lot of memory, making calculations slow or even impossible without access to supercomputing resources.

Progress has been made in both of these fields to overcome their respective weaknesses. Much literature has been published by astronomers on strategies to make the power-spectrum method work when data points are missing \citep[e.g.][]{Stutzki+98, Bensch+01, Ossenkopf+08, Arevalo+12, Benoit-Levy+13, Raghunathan+19}. At the same time, an algorithm has been published to allow for fast computation of semivariograms using the fast Fourier transform \citep{fast_svg}. 

In other words, members of these mathematical communities have developed novel methods for improving all of these techniques in isolation. We advocate for more communication between members of these different communities for cross-disciplinary approaches for data analysis to be shared, so that the lessons learned in one field may benefit researchers in other disciplines.

\section{Afterthoughts}
\label{sec:summary}

In this Note, we've constructed a Rosetta Stone for quantifying spatial correlation. We journeyed to three different subfields where spatial statistics are used to describe data, and found that the techniques developed on these different islands are really not as different as we had thought when we set off. The approaches of the cosmologist, the geostatistician, and the turbulence researcher appear quite different at first glance, but their respective tools, the power spectra, semivariograms, and structure functions, actually capture exactly the same information about how random fields behave. We show in the infographic displayed in Figure \ref{fig:methods_flow} how these are all connected to the fundamentals that we defined in Section \ref{sec:classical_stats}, and their natural extensions for spatial data (Section \ref{sec:spatial_stats}).

We began this Note with an observation: \textit{Things that are close to each other tend to be similar in other ways}. As we finish this piece, we realise that the very reason it was written is because the two authors are both PhD students at the same university with desks fifteen paces apart from each other. We found ourselves facing the same problems of quantifying spatial correlations in astronomical data, but with very different tools in our hands.

In science, we often find ourselves isolated on our respective islands, unable to communicate with researchers outside of our own particular subfield. Prior to writing this Note, SB's conversations around quantifying spatial correlations were nearly exclusively with cosmologists, but she had a strong intuition that semivariograms seemed to be very similar to (and maybe even the same as) the power spectrum that is used to study matter density fields in the early Universe. Similarly, BM was unable to make sense of the terms used by turbulence researchers, limiting his ability to see how the data that he was analysing with geostatistics could be connected to astrophysical theory about the structure of the chaotic, magnetic interstellar media of galaxies.

We leave the reader with some strong encouragement to talk to other scientists outside your subfields. That could be at colloquia, conferences, university hallways, on the internet, or wherever else good science is done. Yes, the languages spoken in each subfield contain exotic terminology and definitions that will disagree with what you are familiar with, but the payoffs are absolutely worth it. These other islands are not abandoned; they are filled with very smart people who have spent decades working hard to solve the same problems that you have. \textbf{Talk to each other!}

When we zoom out through discussions with other scientists, we can start to orient ourselves. In this Note, we aimed to produce a map of the archipelago of domains where spatial statistics is studied. Instead, we discovered that there are land bridges between all of these different islands, and all of us are on a collective scientific Pangea. We have taken you on a journey with us, only to show you that we never really left home. To help you on your own journeys, we leave you with this glossary.

\section*{Glossary of terms} 
\label{sec:glossary}

\textbf{angular power spectrum} -- the angular analog to the \textit{power spectrum} defined in Section \ref{ssec:cosmo}. In the derivation of the $k$-mode power spectrum, functions are imagined to be sums of many waves in space. Similarly, in the derivation of the angular power spectrum, functions are instead defined in terms of \textit{spherical harmonics}. The spherical harmonics represent the fundamental modes of ``vibration" on a sphere. The angular power spectrum is the classic way to quantify spatial correlation in the cosmic microwave background (CMB) because it is almost entirely \textit{Gaussian}.

\textbf{autocorrelation} -- the \textit{cross-correlation} of a signal with a copy of itself that has been shifted in space or time. Unfortunately, ``cross-correlation" has many different definitions that are all in use (Equations \ref{eq:xcorr_def1}, \ref{eq:xcorr_def2}, and \ref{eq:xcorr_def3}), so this term can be a bit ambiguous to use.

\textbf{autocovariance} -- the \textit{covariance} of a signal with a copy of itself that has been shifted in space or time. It is equivalent to the \textit{cross-covariance} (Equation \ref{eq:xcov}) of a signal with itself.

\textbf{Bessel's correction} -- the use of $n-1$ in the denominator in the equation for variance (Equation \ref{eq:var}) rather than $n$, so that the sample variance is an unbiased estimator of the true variance. 

\textbf{bispectrum} -- similar to a \textit{power spectrum}, but using the third cumulant of a \textit{random field} instead of its variance. Explaining what a ``third cumulant" is goes beyond the scope of this work -- but just know that it can be used to show how different a random field is from a \textit{Gaussian} one.

\textbf{central limit theorem} -- see \textit{Gaussian}.

\textbf{configuration space} -- a cosmologist's term for space before any Fourier transforms are applied. In other words, configuration space is simply real space. We use this term because if we just called it ``space" people might think we are talking about Fourier space.

\textbf{convolution} -- given two \textit{random fields} $Z_1(\vec{x})$ and $Z_2(\vec{x})$, the convolution of $Z_1$ and $Z_2$ is given by $Z_1 * Z_2 (\vec{x}) = \sum_{\vec{y}}Z_1(\vec{y})Z_2(\vec{x} - \vec{y})$, where the summand is taken over all pairs for which both $Z_1(\vec{y})$ and $Z_2(\vec{x} - \vec{y})$ are defined. This function sees a lot of use in the field of signal processing. The \textit{Fourier transform} of the convolution of two random fields is equivalent to the pointwise product of the two random fields in Fourier space ($ \mathcal{F}\{Z_1 * Z_2\}  = \mathcal{F}\{Z_1 \} \mathcal{F}\{ Z_2\}$). This equation is often used to quickly calculate convolutions via the fast Fourier transform algorithm.

\textbf{convolutional neural network (CNN)} -- a kind of neural network in which repeated convolutions are performed on the input data. Fortunately, these are well beyond the scope of this work.

\textbf{correlation} -- a statistical measure of the way that two variables are related. If two variables are positively correlated, then if one is measured to be higher than normal, the other one will probably be higher than normal, too. If two variables are negatively correlated, then if one is measured to be higher than normal, the other one will probably be lower than it usually is. Variables with a correlation of zero are \textit{uncorrelated}.\footnote{Note that independence is a stronger condition than being uncorrelated. Independence means there is absolutely no dependence between two random variables. Being uncorrelated only means that there is not any \textit{linear} dependence.} In this Note, we use Pearson's correlation coefficient (Equation \ref{eq:corr}) as our preferred correlation statistic of choice -- anytime we refer to correlation, we refer to this.

\textbf{correlogram} -- another term for the autocorrelation function \citep{Schabenberger+Gotway}.

\textbf{covariance} -- given two random variables $X$ and $Y$, the covariance is a statistical measurement that tells you how changing one affects the other (in other words, it tells you how $X$ and $Y$ \textit{covary}). A positive covariance means that when $X$ is higher than its mean, $Y$ will also be higher than its mean (and when $X$ is lower $Y$ will be lower). A negative covariance means that when $X$ is higher, $Y$ will be lower (and when $X$ is lower, $Y$ will be higher). A mathematical definition for covariance is given in Equation \ref{eq:cov}.

\textbf{cross-correlation} -- unfortunately, many different mathematical functions all share the name of cross-correlation (Equations \ref{eq:xcorr_def1}, \ref{eq:xcorr_def2}, and \ref{eq:xcorr_def3}). For this reason, when reading about cross-correlations or autocorrelations, it is important to pay attention to which definition the author has chosen to use, because it may not match the definition that you are familiar with. That being said, the intuitive idea behind cross-correlations is the same for all definitions: Take two \textit{random fields} $Z_1(\vec{x})$ and $Z_2(\vec{x})$, shift the second one by a \textit{lag} $\vec{r}$, multiply them together, and take some kind of an average. This tells you how similar values of $Z_1(\vec{x})$ are to values of $Z_2(\vec{x}+\vec{r})$, and is very useful for signal processing. For example, cross-correlation of voltages in time underpins \href{https://www.cv.nrao.edu/~sransom/web/Ch3.html#S7}{radio interferometry} which allows us the most precise localisations of objects in the universe using radio telescopes separated by long distances. 

\textbf{cross-covariance} -- the covariance between one \textit{random field} $Z_1(\vec{x})$ at one position $\vec{x}$, and a second random field $Z_2(\vec{x}+\vec{r})$ at a different location $\vec{x}+\vec{r}$. A definition for this function is given in Equation \ref{eq:xcov}.

\textbf{cumulative distribution function (CDF)} - the CDF contains the same information as the probability density function described in Section \ref{sec:classical_stats}, except it tells you what the probability is that X will take a value that is equal or less than x: $\text{CDF}(x) = {\rm Pr}(X \leq x)$. We can translate from a probability distribution to a cumulative distribution with integration: $\text{CDF}(x) = \int_{-\infty}^x p(x^{'}) dx^{'}$.

\textbf{cross-power spectra} - the Fourier transform of the covariance between two different fields or time series. For example, $P_{xy}(k) \equiv \int^{\infty}_{-\infty} dr e^{-i k r} \langle T_{x}(r) T_{y}(r)\rangle$ is the cross power-spectrum between the field $T_x$, and the field $T_y$, assuming translational invariance (to the second order) and isotropy hold in both fields.

\textbf{distribution} -- in our context, it's a shorter way of saying \textit{probability distribution}.

\textbf{energy spectrum} -- a synonym for \textit{power spectrum} commonly used in the study of turbulence.

\textbf{expected value} -- the value that a random variable, or a function of a random variable, is expected to take. The expected value of a random variable is its \textit{mean}.

\textbf{estimator} -- a function or rule that maps a collection of samples of a random variable to an estimate of a statistic or parameter. We almost always need to use estimators for statistics rather than the statistics themselves as we are not usually able to calculate these quantities for an entire population exactly, as that would require knowing the population exactly. However, we can always use samples of our random variable to calculate sample estimates. Common estimators include the sample mean and sample variance which estimate the expectation value and variance of a random variable, respectively (Equations \ref{eq:mean} and \ref{eq:var}). In this Note, we denote an estimator with a hat -- for example, $\rm \widehat{Var}(X)$ is an estimator of $\rm Var(X)$.

\textbf{Fourier transform} -- an integral transform which allows us to express our data in terms of frequencies (or scales for our \textit{random fields}) by decomposing a signal into a sum of waves. See Section \ref{ssec:cosmo} for further details.

\textbf{Gaussian} - shorthand for the \textit{Gaussian distribution}\footnote{Named after Carl Freidrich Gauss.}, or \textit{normal distribution}\footnote{Named after Karl Fredrick Normal.}. If a random variable is known to have a Gaussian distribution, then you can describe its entire probability distribution (and therefore know everything about it) simply by knowing its \textit{mean} and \textit{variance}. For a random variable $X$ with a mean of $\mu$ and a variance of $\sigma^2$, the probability of $X$ taking any value $x$ is $P(X=x) = \frac{1}{\sqrt{2\pi\sigma^2}} \exp \left( -\frac12 {\left( \frac{x - \mu}{\sigma} \right)}^2 \right)$. This definition can be extended to the multivariate case. Consider a collection of random variables $X_1, X_2, \dots, X_n$ (which we can stack into a one-dimensional column matrix called $\mathbf{X}$). If we know that each of these distributions has a mean of $\mu_1, \mu_2, \dots, \mu_n$ (which we stack into a matrix called  $\mathbf{\mu}$), and we have the covariance matrix between all of these random variables (we call this $\mathbf{\Sigma}$, where $\mathbf{\Sigma}_{ij} = \text{Cov}(X_i, X_j)$), then we again have enough information to completely know the probability distribution of (and therefore, everything we could possibly want to know about) this collection of random variables -- it is $P(\mathbf{X}=\mathbf{x}) = \frac{1}{\sqrt{(2\pi)^n|\mathbf{\Sigma}|}} \exp \left( -\frac12 {\left(\mathbf{x} - \mathbf{\mu} \right)}^T \mathbf{\Sigma}^{-1} \left(\mathbf{x} - \mathbf{\mu} \right) \right)$. Here, $|\mathbf{\Sigma}|$ is the \textit{determinant} of the covariance matrix, which roughly speaking tells you about the overall \textit{size} of the variance in your data in the same way that $\sigma^2$ does for the one-dimensional case. The normal distribution is important for statisticians because it comes up all the time -- literally. There's a theorem in mathematics (the \textit{central limit theorem}\footnote{A very nifty and well-explained proof of the central limit theorem can be found in \href{https://www.cs.toronto.edu/~yuvalf/CLT.pdf}{this note}.}) that states that if you take enough independent observations of \textit{any} random variable and average them, then after infinite observations, the distribution you get will \textit{always} be \textit{exactly} the normal distribution, irrespective of the distribution that you started with. Of course, taking infinite samples of a random variable is infinitely expensive and takes infinite time, so we often have to make models of our random variables that take into account \textit{non-Gaussianities} -- that is, differences between the actual distributions that we see and the simple, idealised normal distributions.
  
\textbf{homogeneous} -- a \textit{random field} is homogeneous if it is the same at all points. Exactly \textit{what } needs to be the same at each point is not specified by this term. Sometimes it is used to mean that the exact values of the random field must be the same everywhere. Other times it is used to mean that the second-order or higher-order statistics of a random field must be the same everywhere -- see \textit{stationary}.

\textbf{isotropic} -- a \textit{random field} is isotropic if it looks about the same in every direction (i.e. there is \textit{no preferred direction}). In other words, it looks statistically the same even if it is rotated (i.e. it is \textit{rotationally invariant}).

\textbf{k-mode} -- the way cosmologists define their Fourier space modes, $k = \frac{2 \pi}{r}$, where r is a distance. In the temporal domain, we often describe our modes as frequencies, and k-modes are spatial analogs that can be thought of as spatial frequencies. Here we show the a k-mode in 1D, but we can easily extend this to 3D with $\mathbf{k} = (k_x, k_y, k_z) = \left(\frac{2 \pi}{r_x}, \frac{2 \pi}{r_y}, \frac{2 \pi}{r_z}\right)$.

\textbf{l-mode and m-modes} -- the angular analogs to the k-mode that are indices of the spherical harmonics that allow us to expand functions on a sphere.

\textbf{lag} -- this is a word that geostatisticians and signal processing people sometimes use, but it just means ``separation". In time series data, temporal lag refers to the separation in time between two signals: Given two events at $Z(t_1)$ and $Z(t_2)$, the lag between them is $t_2 - t_1$. This distance is often given the symbol $\tau$. In spatial data, spatial lag refers to the separation in space between two signals: Given two events at $Z(\vec{x}_1)$ and $Z(\vec{x}_2)$, the lag between them is $\vec{x}_2 - \vec{x}_1$. In this Note, we use the symbol $\vec{r}$ to refer to this separation.  

\textbf{mean} -- a measure of the centre of a random variable. It is our best guess for the approximate value that this random variable should take. For this reason, it is also called the \textit{expected value} (Equation \ref{eq:mean}).

\textbf{median} -- an alternative measure of the centre of a random variable. It is the most central value of the random variable that we measure -- if we measured a random variable 101 times, 50 of our samples will be below the median, 50 will be above the median, and one (the middle one) will be the median.

\textbf{mode} -- another alternative measure of the centre of a random variable. It is the most likely value for our random variable to be, or the value that appears most often in our sample. For a \textit{Gaussian} random variable, the mean, median, and mode will all be equivalent -- but this should not be expected to be true for other kinds of distributions.

\textbf{moment} -- the $n$-th moment of a random variable $X$ is the expected value of $X^n$. The first moment of a random variable is its \textit{mean}. The second moment of a random variable is related to its \textit{variance}. Higher order moments tell you about other properties of the distribution of a random variable that we do not cover in this Note. 

\textbf{periodogram} -- a common way of estimating the power spectrum of time series data. The power spectrum and periodogram are exactly equivalent as described in \citet{VanderPlas_2018}, except the periodogram is commonly divided by the number of frequencies. It's optimal for estimating the power spectrum for time series data that have been unevenly sampled. If you want to learn more about the periodogram, \citet{VanderPlas_2018} provides an intuitive introduction to this topic and more generally quantifying correlations in time series data.

\textbf{p-value} -- a kind of calculation in which we assume something which we think is wrong (the \textit{null hypothesis}), and then compute the probability of that wrong thing being right. If we measure a low probability (a low \textit{p-value}) of the thing we think is wrong being correct, then we can conclude, with some confidence, that the thing we thought was wrong might actually be wrong (we can \textit{reject the null hypothesis}). The problem is, such an analysis won't always tell you what the right thing to believe is instead.

\textbf{power spectral density (PSD)} -- see \textit{power spectrum}. This term seems to be more commonly used in more engineering related fields, and is generally used in reference to time-series data. While cosmologists talk about power spectra, the engineers that build the telescopes that cosmologists use to capture the power spectra will call this same statistical measure the power spectral density. Astronomers who study pulsar timing arrays also prefer the term \textit{power spectral density}.

\textbf{power spectrum} - the Fourier transform of the covariance function (Equation \ref{eq_matter_pspec_full}). Many cosmology textbooks will define the power spectrum as the Fourier transform of the \textit{two-point correlation function} instead -- but beware! The two-point correlation function has multiple definitions, not all of which are equivalent, and the standard definition of the two-point correlation function (Equation \ref{eq:2pcf}) can only be used to define the power spectrum if the \textit{random field} under investigation has a mean of zero everywhere.

\textbf{probability distribution} -- a function that completely describes a random variable. For every value that the random variable can possibly take, the probability distribution tells you how likely each possible value is to occur. 

\textbf{random variable} -- a mathematical structure that is used to describe something whose value is not certain. Different measurements of a random variable may result in different values. A random variable can be fully described by its \textit{probability distribution}.

\textbf{random field} -- a mathematical structure that is used to describe a system that is stochastic in space or time. In this structure, every point in the domain can be described by a \textit{random variable}. In general, points that are closer to each other tend to be more tightly \textit{correlated} in their values -- this is Tobler's First Law of Geography. In this Note, we provide explanations of many different techniques that are used to analyse the correlated spatial structure of random fields.

\textbf{second-order stationary} -- a \textit{random field} is second-order \textit{stationary} if it follows Condition \ref{eq:weak-sense-stationary} -- that is, if it has a constant \textit{mean} throughout, and the \textit{covariance} between the values of the random field at any two points $\vec{x}$ and $\vec{y}$ depends only on the separation between $\vec{x}$ and $\vec{y}$. If a field satisfies this second condition, then it will also necessarily have a constant \textit{variance} throughout (to use a statistics term, it will be homoskedastic). If a field is second-order stationary, then we can define its second-order statistics (its covariance and \textit{correlation}) as functions of one variable: the separation between data points.

\textbf{semivariogram} -- a statistical tool from the realm of geostatistics that is useful for visualising how the \textit{variance} between data points increases with their separation. We give a mathematical definition in Equation \ref{eq:svg}. For second-order \textit{stationary} \textit{random fields} (where the \textit{covariance} between data points depends only on their separation), the semivariogram is closely related to the covariance function, as shown in Equation \ref{eq:svg_to_2pc}.

\textbf{serial variation function} -- an old historic word for the \textit{semivariogram}. The plot of this function was also called the \textit{serial variation curve} \citep{Jowett+55b, Jowett+55a}.

\textbf{signal to noise ratio (SNR)} --  in astronomy and beyond, a very useful measure for how well you can see something is the \textit{signal to noise ratio}, often abbreviated to SNR, or S/N. This is the intensity (it could be brightness, or loudness, or spectral flux density) of a signal, divided by the \textit{standard deviation} (or the noise level, $\sigma$) of that signal. Because the standard deviation has the same units as the signal itself, the SNR will always be dimensionless, regardless of what we are measuring or what units we use to measure it. You may not understand what magnitudes or decibels or janskies are, but you can always understand what a SNR of 10 means: the signal is ten times stronger than our uncertainty about it. This is also called a 10$\sigma$ detection. 

\textbf{standard deviation} -- a useful metric that captures how much a \textit{ random variable} tends to vary. It is equal to the square root of the \textit{variance}, and is usually denoted by the symbol $\sigma$. The units of the standard deviation will always be the same as the units of the original random variable.

\textbf{stationary} - originally used with time series data, this word is a synonym for \textit{homogeneous}. Just like the word \textit{homogeneous}, 
the term stationary is used to refer to two different conditions -- see the entries for \textit{strictly stationary} and \textit{second-order stationary}.

\textbf{strictly stationary} -- a \textit{random field} is strictly \textit{stationary} if \textit{all} of its statistical properties are translationally invariant. This is stronger than the assumption of second-order stationarity because it also assumes that all higher-order statistics of the data field do not depend on the absolute position of data points, and only on their separation relative to each other.

\textbf{strongly stationary} -- a synonym for \textit{strictly stationary}.

\textbf{translational invariance} -- see \textit{homogeneous}.

\textbf{trispectrum} -- the Fourier transform of the fourth cumulant of a \textit{random field}. Like the \textit{bispectrum}, it can also be used to search for and characterise non-Gaussianities. Also, like the bispectrum, describing it properly in a way that would do it justice is beyond the scope of this Note -- see \citet{Sefusatti_2005} which describes a trispectrum estimator.

\textbf{two-point correlation function} - a way to quantify the clustering of values in a \textit{random field} as a function of separation, commonly applied in cosmology. One common definition is that it is the function that describes how likely it is to find pairs of points at a given distance compared to a random distribution (Equation \ref{eq:prob_def_of_2pc}). The two-point correlation function can be viewed as the continuous analogue of the covariance matrix. However, as described in the Boxes on pages 12–14, there are multiple definitions of the correlation function used in astrophysics, many of which overlap with what is sometimes referred to as the autocorrelation function. These definitions do not always align with the variance normalized version preferred by statisticians presented in Equation \ref{eq:corr_fn}. Confusingly, it is not a correlation coefficient -- it has units equal to the units of the random field squared, and it is not normalised to lie between $-1$ and $1$.

\textbf{variance} -- a statistical measure of how much a random variable tends to \textit{vary}. It is computed by summing the squared difference between every data point and the \textit{mean}, and dividing by the number of data points minus one (Equation \ref{eq:var}). The units of variance will be equal to the square of the units of the random variable. The \textit{covariance} between a variable and itself is its variance.

\textbf{variogram} -- is a tricky word. In most instances, it is a synonym for \textit{semivariogram} -- this is how Georges Matheron originally used it, and how it is used in some textbooks \citep[e.g.][]{chiles_delfiner99}. Some other
authors define it differently, saying that the variogram is equal to twice the semivariogram. For example, \citet{Schabenberger+Gotway} call out \citet{chiles_delfiner99} by name for their definition of a variogram, stating that ``there is nothing ``established" about being off the mark by factor 2", and that the semivariogram as we defined it shouldn't be called a variogram because ``the savings in ink are disproportionate to the confusion created when “semi” is dropped". We recommend simply avoiding the use of the term variogram entirely -- since it usually (but not always) means the same thing as semivariogram, it's much less confusing to only talk about semivariograms. 

\textbf{wavenumber} -- a synonym for \textit{k-mode}, either a vector or its magnitude.

\textbf{weakly stationary} -- a synonym for \textit{second-order stationary}.s

\textbf{weak-sense stationary} -- a second synonym for \textit{second-order stationary}.

\textbf{Wiener-Khinchin Theorem (also Wiener-Khintchine Theorem)} -- this allows us to relate the \textit{two-point correlation function} and the \textit{power spectrum} through the Fourier transform. It shows that the two contain exactly the same information. Although we do not discuss it in depth in this note, it's essential in allowing the spectral decomposition of the two-point correlation function. Find its proof \href{https://mathworld.wolfram.com/Wiener-KhinchinTheorem.html}{here} or in \citet{2017mcp..book.....T}.

\textbf{wide-sense stationary} -- a third synonym for \textit{second-order stationary}.

\textbf{window function} -- 
a window function is a weighting function that describes how the values of a \textit{random field} change due to forces other than spatial clustering, such as foregrounds, instrumental effects, and survey geometry \citep{Karim+23}. \citet{Gorce_2023} shows how window functions affect cosmological measurements. Estimating these functions is crucial in real-world measurements of the \textit{power spectrum}.

\section*{Acknowledgements}
B.M. and S.B. contributed to all roles in writing this Note. We also outline the roles that each of the authors \textbf{led} in the creation of this Note according to the Contributor Roles Taxonomy
(CRediT)\footnote{https://credit.niso.org/}. B.M. led Writing – original draft, Data curation, and Methodology. S.B. led Software and Resources, and both authors contributed equally to Visualization.

We thank all of our friends and colleagues for their fruitful conversations about this Note, including Jose Fuentes Baeza, Aman Chokshi, Justin Clancy, Nicolò Dalmasso, Dillon Dong, Adélie Gorce, Bradley Greig, Lisa McBride, Shona McEvoy, Kevin Levy, Robert Pascua, Christian Reichardt, Michele Trenti, and Stuart Wyithe. We also thank Tingjin Chu, Alex Clark, Veronica Dike, Nicholas Rui, Amit Seta, and Neco Kriel for providing excellent written feedback to specific sections of this Note. We would further like to thank the anonymous referee, whose careful reading of this manuscript improved the quality of this work.

A special thank you goes to Tree Smith for giving the Masters' talk that kickstarted the two-hour argument between us which inspired us to write this Note. 

BM thanks Alex Clark for math inspiration. 

BM and SB are also incredibly thankful to Tong Cheunchitra for providing thorough comments on this Note that gave us new insights we hadn't thought of.

SB thanks Adrian Liu for a set of \href{https://github.com/acliu/21cmStarterKit/blob/57e1af5ca9c0b8b098ca35942278439daaf94a72/Basics/PowerSpectra.pdf}{notes} on the power spectra that were useful in its definition and comparison to other methods. 

BM acknowledges support from Australian Government Research Training Program (RTP) Scholarships and The David Lachlan Hay Memorial Fund. SB is supported by the Melbourne Research Scholarship and N D Goldsworthy Scholarship for Physics. This research is supported in part by the Australian Research Council Centre of Excellence for All Sky Astrophysics in 3 Dimensions (ASTRO 3D), through project number CE170100013. The majority of this research was conducted on Wurundjeri, Ngunnawal (Ngunawal), and Ngambri land. Sovereignty was never ceded.

\bibliography{main}{}
\bibliographystyle{aasjournal}
\end{document}